\documentclass[USenglish,onecolumn]{article}

\usepackage[utf8]{inputenc}
\usepackage{comment}
\usepackage[big]{dgruyter}
\usepackage{booktabs}
\usepackage{multirow}
\usepackage{rotating}
\usepackage{threeparttable}
\usepackage{array}
\usepackage{graphicx}
\usepackage{subcaption}
\usepackage{algorithm}
\usepackage{xcolor}
\usepackage{float}
\makeatletter
\def\plist@algorithm{Alg.\space}
\makeatother

\begin{document}

\articletype{Research Article{\hfill}Open Access}

  \author[1]{Jing Li}
   \affil[1]{CAUSALab, Department of Epidemiology, Harvard T.H Chan School of Public Health, Boston, MA, USA; Corresponding Author; E-mail: jing\_li@hsph.harvard.edu}

\author[2]{Sophia M Rein}
  \affil[2]{CAUSALab, Department of Epidemiology, Harvard T.H Chan School of Public Health, Boston, MA, USA; E-mail: srein@hsph.harvard.edu}

  \author[3]{Zilong Wang}
  \affil[3]{Department of Biostatistics, Harvard T.H Chan School of Public Health, Boston, MA, USA; E-mail: zilongwang@hsph.harvard.edu}

  \author[4]{Zixiao Wang}
  \affil[4]{CAUSALab, Department of Epidemiology, Harvard T.H Chan School of Public Health, Boston, MA, USA; E-mail: zixiaowang@fas.harvard.edu}

  \author[5]{Sean McGrath}
  \affil[5]{Department of Statistics and Data Science; Bursky School of Public Health, Washington University in St. Louis, St. Louis, MO, USA; E-mail: mcgrathsean@wustl.edu}

  \author[6]{James M. Robins}
  \affil[6]{CAUSALab, Departments of Epidemiology and Biostatistics, Harvard T.H Chan School of Public Health, Boston, MA, USA; E-mail: robins@hsph.harvard.edu}

  \author[8]{Thomas Kolokotrones}
  \affil[8]{Department of Epidemiology, Harvard T.H Chan School of Public Health, Boston, MA, USA; E-mail: tkolokot@hsph.harvard.edu}

  \author[7]{Rajarshi Mukherjee}
  \affil[7]{Department of Biostatistics, Harvard T.H Chan School of Public Health, Boston, MA, USA; E-mail: ram521@mail.harvard.edu}

  \author[9]{Miguel A. Hern\'{a}n}
  \affil[9]{CAUSALab, Departments of Epidemiology and Biostatistics, Harvard T.H Chan School of Public Health, Boston, MA, USA; E-mail: mhernan@hsph.harvard.edu}

\title{\huge A plug-in estimator of the non-iterative conditional expectation g-formula using long short-term memory models}

\runningauthor{J. Li et al.}
\runningtitle{A Plug-in Estimator of the NICE G-Formula Using LSTMs}

\begin{abstract}
{The non-iterative conditional expectation (NICE) plug-in estimator of the g-formula can be used to estimate the causal effects of sustained treatment strategies, even in the presence of treatment-confounder feedback, under the assumptions of exchangeability, positivity, and consistency. In practical applications, parametric models are commonly used to estimate the joint distribution of confounders, treatments, and outcomes over time.  
As an alternative to parametric models, long short-term memory (LSTM) models, designed to capture both short-term patterns and long-range dependencies in sequential data, may better accommodate complex temporal dependencies among time-varying confounders, treatments, and outcomes. However, the validity of statistical inferences of a plug-in estimator based on LSTM models has not yet been investigated. Here, we provide proof of principle that, under a complex data generation mechanism that resembles data from a cohort of persons with HIV and sufficiently large sample sizes, an LSTM-based NICE g-formula plug-in estimator results in less bias than parametric estimators while achieving near-nominal coverage for 95\% confidence intervals. This work is intended to spur the development of a theoretical framework that encompasses deep learning plug-in estimators for causal inference based on high-dimensional longitudinal data.}
\end{abstract}

\keywords{causal inference; g-formula; uncertainty quantification; long short-term memory; longitudinal data}
\classification[MSC]{62D20, 62M45, 62G09, 62G15}



\maketitle
\section{Introduction}

The g-formula identifies the causal effects of sustained treatment strategies from observational data under the assumptions of consistency, positivity, and exchangeability \cite{ROBINS19861393}. 
A common representation of the g-formula is a non-iterative conditional expectation (NICE) of the outcome weighted by the joint density of covariate history. 
When covariates are high-dimensional, nonparametric estimation of the components of the NICE g-formula--the conditional distributions of confounders, treatments and outcome--is generally infeasible. Instead, they must be estimated using non-saturated models and plugged into the NICE representation. The resulting integral is then approximated using Monte Carlo simulation.

In previous applications, the components of the NICE g-formula have been typically estimated using parametric models, such as generalized linear models \cite{taubman2009intervening, lajous2013changes, danaei2013hypothetical, garcia2014incidence, Lodi, zhang2018comparing, Dickerman, Jain,10.1093/aje/kwad143}.
Because the validity of the inference relies on the correct specification of such models~\cite{MCGRATH2020100008}, it would be advantageous to replace the parametric models by more flexible approaches. For example, recurrent neural networks can handle complex long-range dependencies among time-varying confounders, treatments, and outcome and do not require parametrically-constrained functional forms for all variables, thereby reducing the risk of model misspecification. 

However, existing statistical theory provides little guidance about the validity of this "deep learning" approach to the plug-in estimator of the NICE g-formula. Standard asymptotic results on plug-in estimators of the g-formula do not directly apply when using neural networks because they typically assume low-dimensional models for the nuisance functions. Also, the flexibility of neural networks may be effective for bias reduction, but it makes their asymptotic behavior depend on numerous factors, including network architecture, training algorithms, and regularization choices. The need to estimate high-dimensional, time-varying nuisance functions further complicates the theoretical analysis.
 
 Moreover, the practical implementation of uncertainty quantification via resampling-based approaches, such as the bootstrap, is computationally intensive, requiring repeated training of multiple neural networks for each resampled dataset. These high computational requirements also make it difficult to evaluate the validity of the statistical inferences. In fact, simulations have been used to demonstrate the lower bias of a plug-in NICE g-formula estimator using long short-term memory (LSTM) neural networks  \cite{rein2024deep} compared with parametric models, but not to evaluate the validity of the statistical inferences.

 In this paper, we provide proof of concept for the validity of an LSTM-based estimator of the NICE g-formula. Our estimation procedure incorporates sample splitting and cross-fitting \cite{newey2018cross,chernozhukov2018double,liu2020nearly, mcgrath2022undersmoothing} to generate estimates of the LSTM-based estimator of the NICE g-formula and 95\% confidence intervals around the estimates. Our goal is to assess, in complex and realistic simulations, both the bias reduction and the validity of statistical inferences when the NICE g-formula is combined with deep learning. This work is intended to spur the development of a theoretical framework that encompasses deep learning plug-in estimators for causal inference based on high-dimensional longitudinal data.

The paper is structured as follows. In Section \ref{subsec:g-formula}, we introduce the g-formula for time-varying treatments and a failure time outcome, as well as its NICE plug-in estimator with parametric models. In Section \ref{subsec:lstm_estimator} and \ref{subsec: sample_splitting}, we describe the modifications required to replace parametric models with LSTM models and introduce an approach based on sample splitting and cross-fitting. In Section \ref{subsec:simulation} and \ref{subsec:analysis}, we describe simulations under both simple and complex longitudinal data-generating mechanisms, assess finite-sample performance (in terms of bias, variance and coverage of nominal 95\% confidence intervals), and compare the proposed estimator with the parametric NICE g-formula under different sample sizes. In Section \ref{sec:results}, we present the results. We end with a discussion in Section \ref{sec:discussion}.

\section{Methodology}

\subsection{The NICE g-formula}
\label{subsec:g-formula}

We begin by introducing notation for longitudinal studies with time-varying treatments and a failure time outcome. Let $k = 0, 1, 2, ..., K$ represent the time intervals (e.g., months) of follow-up. Let $L_k$ denote a vector of observed time-varying confounders measured at time $k$, $A_k$ an indicator of treatment at time $k$, and $Y_k$ an indicator of failure (the outcome) by time $k$. By definition, $Y_0 = 0$. We adopt the convention that the order of the observed data is $(L_{k}, A_{k}, Y_{k+1})$. The history of random variable $X$ through $k$ is denoted by $\bar{X}_k=(X_0,...,X_k)$.

A treatment strategy $g=\{g_0,...,g_K\}$ is a rule that assigns treatment at each time $k$ according to an independent draw from an intervention density $f^{\text{int}}(a_k \mid \bar{l}_{k}, \bar{a}_{k-1})$ where $(\bar{l}_{k}, \bar{a}_{k-1})$ is a realization of $(\bar{L}_{k}, \bar{A}_{k-1})$. Our target estimand is 
\begin{equation}
\mathbb{E}\!\bigl[Y^{g}_{K+1}\bigr]
\;=\;
\Pr\!\bigl(Y^{g}_{K+1} = 1\bigr),
\end{equation}

that is, the marginal probability (risk) of $Y$ by time $K+1$ under treatment strategy $g$.

For a vector of discrete variables $L_k$ and $A_k$, the risk of the outcome at $K+1$ under an intervention density $f^{int}(a_k|\bar{l}_{k},\bar{a}_{k-1})$ is identified by the g-formula:

\begin{align*}
& \sum_{\bar{a}_K} \sum_{\bar{l}_K} \sum_{k=0}^{K} 
\Pr\left(Y_{k+1} = 1 \mid \bar{L}_k = \bar{l}_k, \bar{A}_k = \bar{a}_k, \bar{Y}_k = 0\right) \times \\
& \quad \prod_{j=0}^{k}
\Big\{ 
f\left(l_j \mid \bar{l}_{j-1}, \bar{a}_{j-1}\right) 
f^{\text{int}}\left(a_j \mid \bar{l}_j, \bar{a}_{j-1}\right) 
\times \\
& \quad \Pr\left(Y_j = 0 \mid \bar{L}_{j-1} = \bar{l}_{j-1}, \bar{A}_{j-1} = \bar{a}_{j-1},\bar{Y}_{j-1} = 0\right)
\Big\}
\end{align*}

under the assumptions of consistency, positivity, and exchangeability described in Appendix \ref{appendix:identify}. For continuous variables, the g-formula is written in terms of integrals rather than sums.

In most previous applications, each of the conditional densities in the g-formula has been estimated via a parametric model, e.g., a linear regression model for continuous variables, and a logistic regression model for binary variables. The model estimates are then plugged into the g-formula above with treatment values replaced by those values mandated by the intervention density $f^{\text{int}}(a_k \mid \bar{l}_{k}, \bar{a}_{k-1})$.

Because the sum (or integral) over all treatment and confounder histories may be computationally intractable, it is typically approximated via Monte Carlo simulation. Nonparametric bootstrapping is used to construct 95\% confidence intervals around the g-formula estimates, with the standard error estimated as the sample standard deviation of the bootstrap estimates. See \cite{young2011comparative,young2014identification,young2020causal}
for a more detailed description of the g-formula estimation algorithm and extensions to accommodate scenarios with competing risks and interventions that may depend on the natural value of treatment.  

\subsection{NICE g-formula estimation via LSTM models}
\label{subsec:lstm_estimator}

As proposed by \cite{rein2024deep}, the parametric models in the above plug-in estimator can be replaced by two LSTM models \cite{hochreiter1997long}. The first LSTM model is used to estimate the joint distribution of all covariates (confounders and treatments) conditional on past histories:

\begin{equation}
\begin{aligned}
\prod_{k=0}^{K} f(l_k, a_k \mid \bar{l}_{k-1}, \bar{a}_{k-1})
&= \prod_{k=0}^{K} f(l_k \mid \bar{l}_{k-1}, \bar{a}_{k-1}) \times f^{\text{obs}}(a_k \mid \bar{l}_{k}, \bar{a}_{k-1})
\end{aligned}
\end{equation}

The second LSTM model is used to estimate the conditional probability of the outcome given covariate history:

\begin{equation}
\prod_{k=0}^{K} \Pr\left(Y_{k+1} = 0 \mid \bar{L}_{k} = \bar{l}_{k}, \bar{A}_{k} = \bar{a}_{k}, \bar{Y}_{k} = 0\right)
\end{equation}

Each LSTM model has two components: an LSTM encoder, and a multilayer perceptron (MLP) decoder:

\begin{itemize}
    \item LSTM encoder: The input to the encoder is a sequence of the multivariate features observed over time (i.e., the covariate history). This sequence is encoded into a latent representation (hidden state vector) after passing through the LSTM neural network. The architecture of the LSTM encoder is determined by three hyperparameters: $layers\_num$, $hidden\_size$, and $dropout$. The $layers\_num$ specifies the number of stacked recurrent layers (i.e., the depth of the LSTM network). The $hidden\_size$ specifies the dimensionality of the hidden state vector in the LSTM at each time step (i.e., the width of the LSTM layer). The $dropout$ specifies the dropout rate applied between LSTM layers, randomly zeroing out some activations during training.
    
    \item MLP decoder: The MLP decoder consists of two linear layers with ReLU activations followed by an output layer. This design enables the model to learn flexible nonlinear mappings from the latent temporal representation produced by the LSTM to the covariate space. The hidden state from the LSTM encoder is passed through an MLP, which produces a vector of all predicted covariate values. The dimensionality of the final output layer matches the number of covariates being jointly predicted. The MLP decoder is specified by the hyperparameter $feature\_dim$, which specifies the size of the intermediate representations when mapping from the LSTM hidden state to the final covariate prediction. In the LSTM model for estimating the outcome distribution, the MLP decoder produces a probability of the outcome in the output layer. 
    See Appendix \ref{app:implementation} for more implementation details.
\end{itemize}

We search over a set of candidate hyperparameter configurations and select the one that results in the lowest validation loss (as defined in Appendix \ref{app:loss}). For each hyperparameter configuration of the covariate model, the candidate hyperparameters are sampled independently from the specified search spaces. 
Discrete hyperparameters -- including the number of recurrent layers, the hidden dimension, the feature dimension, and the batch size -- are drawn uniformly from their 
respective sets. Continuous hyperparameters were sampled according to their 
specified distributions: dropout values are drawn from a uniform distribution, learning rates were sampled from a log-uniform distribution, 
and weight decay parameters are sampled from a 
log-uniform distribution. The full hyperparameter list and their ranges are provided in Appendix~\ref{app:space}. 

We then split the dataset into a training set and a validation set. The LSTM models are trained on the training set and evaluated on the validation set.
For each hyperparameter configuration, the LSTM model is trained for up to 10,000 epochs to minimize the training loss. When there is no decrease in validation loss for 50 consecutive epochs, model training was stopped, and the lowest validation loss observed during training was recorded for that configuration. The final configuration was selected as the successful trial with the lowest validation loss across all hyperparameter trials. Hyperparameter tuning was conducted using \texttt{hyperopt}~\cite{bergstra2013tpe} with the tree-structured Parzen estimator algorithm. Algorithms \ref{alg:hyperparameter_covariate} and \ref{alg:hyperparameter_outcome} outline the procedures for hyperparameter optimization for the covariate model and the outcome model, respectively. 

After selecting the hyperparameter values, the LSTM models can be used to estimate the components of the NICE g-formula, which are then plugged into the g-formula. Same as when using parametric models, the sum over all covariate histories can be approximated via Monte Carlo simulation, and the standard error of the estimates can be estimated via nonparametric bootstrapping. However, there is no theoretical guarantee that this procedure would produce valid 95\% confidence intervals. In the next section, we describe how we incorporated sample splitting and cross-fitting in an attempt to improve its inferential properties.

\begin{algorithm*}[t]
\caption{Hyperparameter optimization for covariate model}
\label{alg:hyperparameter_covariate}

\noindent\textbf{Requires:}
Input dataset $\mathcal{D}_x$,
deep model $f_{cov}$,
number of hyperparameter search trials $M$

\par
\noindent\textbf{Step 1: Dataset preparation}

\noindent Split the dataset $\mathcal{D}_x$ into training set
$\mathcal{D}_{train}$ and validation set $\mathcal{D}_{val}$
with a ratio of 4:1.

\noindent Standardize the covariate values in the training dataset
$\mathcal{D}_{train}$ and validation dataset $\mathcal{D}_{val}$.

\par
\noindent\textbf{Step 2: Define the search space}

\noindent For each hyperparameter in the LSTM model $f_{cov}$,
specify the search space (i.e., the range of values to be explored).

\par
\noindent\textbf{Step 3: Hyperparameter tuning}

\noindent\textbf{for} $m=1$ to $M$ \textbf{do}

\begin{list}{}{\leftmargin=2em \rightmargin=0pt
                \topsep=0pt \itemsep=0pt \parsep=0pt}
\item[] Train the LSTM model $f_{cov}$ on the training dataset
$\mathcal{D}_{train}$.

\item[] Evaluate the model on the hold-out validation dataset
$\mathcal{D}_{val}$, calculate the average loss across all individuals
and all time points for each covariate, and then sum the losses over
all covariates.
\end{list}

\noindent\textbf{end for}

\par
\noindent\textbf{Step 4: Select the best hyperparameter configuration}

\noindent Select the hyperparameter set for model $f_{cov}$ that
achieves the lowest validation loss.

\par
\noindent\textbf{Return}
A hyperparameter set for $f_{cov}$.

\end{algorithm*}

\begin{algorithm*}[t]
\caption{Hyperparameter optimization for outcome model}
\label{alg:hyperparameter_outcome}

\noindent\textbf{Requires:}
Input dataset $\mathcal{D}_x$,
deep model $f_{out}$,
number of hyperparameter search trials $M$

\par
\noindent\textbf{Step 1: Dataset preparation}

\noindent Split the dataset $\mathcal{D}_x$ into training set
$\mathcal{D}_{train}$ and validation set $\mathcal{D}_{val}$
with a ratio of 4:1.

\noindent Standardize the covariate values in the training dataset
$\mathcal{D}_{train}$ and validation dataset $\mathcal{D}_{val}$.

\par
\noindent\textbf{Step 2: Define the search space}

\noindent For each hyperparameter in the LSTM model $f_{out}$,
specify the search space (i.e., the range of values to be explored).

\par
\noindent\textbf{Step 3: Hyperparameter tuning}

\noindent\textbf{for} $m=1$ to $M$ \textbf{do}

\begin{list}{}{\leftmargin=2em \rightmargin=0pt
                \topsep=0pt \itemsep=0pt \parsep=0pt}
\item[] Train the LSTM model $f_{out}$ on the training dataset
$\mathcal{D}_{train}$.

\item[] Evaluate the model on the hold-out validation dataset
$\mathcal{D}_{val}$ and calculate the average outcome loss across
all individuals and all time points.
\end{list}

\noindent\textbf{end for}

\par
\noindent\textbf{Step 4: Select the best hyperparameter configuration}

\noindent Select the hyperparameter set for model $f_{out}$ that
achieves the lowest validation loss.

\par
\noindent\textbf{Return}
A hyperparameter set for $f_{out}$.

\end{algorithm*}

\subsection{Sample splitting and Cross-fitting}
\label{subsec: sample_splitting}

\begin{algorithm*}[!tb]
\caption{Estimation of the LSTM-based NICE g-formula estimator}
\label{alg:sample-splitting}

Dataset $\mathcal{D} = \{(l_i,a_i,y_i)\}_{i=1}^{N}$,
LSTM model $f_{cov}$ for covariates,
LSTM model $f_{out}$ for the outcome,
number of bootstrap samples $B$,
number of hyperparameter search trials $M$.

\par\medskip
\textbf{Step 1: Sample splitting}

Randomly partition the dataset $\mathcal{D}$ into halves
$\mathcal{D}_{1}$ and $\mathcal{D}_{2}$.

\par\medskip
\textbf{Step 2: Hyperparameter tuning on $\mathcal{D}_{1}$}

Conduct an $M$-trial search to select LSTM hyperparameters for both
the covariate and outcome models. An internal training/validation split
within $\mathcal{D}_{1}$ is applied to choose the configuration that
minimizes the validation loss.

\par\medskip
\textbf{Step 3: G-formula estimation on $\mathcal{D}_{2}$}

Train the LSTM models $f_{cov}$ and $f_{out}$ with the chosen hyperparameter configurations on $\mathcal{D}_{2}$, evaluate the trained models on the validation dataset, and select the model that achieves the best performance on the validation criterion.

Use the trained models $f_{cov}$ and $f_{out}$ to estimate the conditional distributions needed for the NICE g-formula.

Perform Monte Carlo simulation by sampling from the estimated distributions that are parameterized by $f_{cov}$ and $f_{out}$.

Calculate the g-formula estimate $\hat{p}_{2}$ under the specified interventions.

\par\medskip
\textbf{Step 4: Bootstrap on $\mathcal{D}_{2}$}

\noindent\textbf{for} $b = 1$ to $B$ \textbf{do}
\begin{itemize}
    \item[] Resample a bootstrap dataset $\mathcal{D}^{(b)}$ from $\mathcal{D}_{2}$
    \item[] Compute the g-formula estimate $\hat{p}^{(b)}$
\end{itemize}

\par\medskip
\textbf{Step 5: Construct confidence interval}

Calculate the estimated standard error $\hat{s}_{2}$ using the bootstrap
estimates $\{\hat{p}^{(1)},\ldots,\hat{p}^{(B)}\}$.

\par\medskip
\textbf{Step 6: Cross-fitting}

Swap the subsamples $\mathcal{D}_{1}$ and $\mathcal{D}_{2}$ and repeat
Steps 2--5.

Obtain the final g-formula risk estimate $\hat{p}$ by averaging
$\hat{p}_{1}$ and $\hat{p}_{2}$.

Compute the final standard error $\hat{s}$ by
\[
\hat{s}
=
\sqrt{\frac{\hat{s}_{1}^{2}+\hat{s}_{2}^{2}}{4}}.
\]

Obtain the confidence interval via normal approximation with standard
error $\hat{s}$.

\par\medskip
\textbf{Return} G-formula estimate $\hat{p}$, standard error $\hat{s}$,
and confidence interval
$\left[\hat{p}^{\mathrm{lb}},\hat{p}^{\mathrm{ub}}\right]$.

\end{algorithm*}

In double machine learning (DML) methods, the construction of estimators requires the derivation of influence functions \citep{chernozhukov2018double,chernozhukov2022locally,kennedy2024semiparametric,westling2024inference}. 
However, deriving influence functions for target parameters and estimating the associated nuisance functions can be challenging, particularly when the nuisance functions are complex, complicating the development of such estimators. Although there exists a dedicated literature attempting to provide a generalized framework for such nuisance estimation, simpler strategies often yield satisfactory performance in specific problems. 
As an example, if one considers the estimation of the average treatment effect and related quantities for time-fixed treatments \citep{crump2009dealing}, one can obtain desired consistent and asymptotically normal (CAN) estimation with plug-in principles. General wisdom dictates that these plug-in estimators may fail in high-dimensional problems, but recent literature has shown otherwise in several cases \citep{yadlowsky2022explaining,mcgrath2022undersmoothing,mcgrath2025optimal,jiang2025new}. It has also been observed \citep{mcgrath2022undersmoothing,mcgrath2025optimal,jiang2025new} that careful sample splitting and cross-fitting can further enhance the benefits of the plug-in principle. Here, we propose a plug-in approach that follows the same philosophy and apply it to our setting, which is substantially more complex than the ones tackled in \citep{yadlowsky2022explaining,mcgrath2022undersmoothing,mcgrath2025optimal,jiang2025new}. A description of the procedure follows.

Before starting the estimation, the dataset $\mathcal{D}$ is randomly split into two halves $\mathcal{D}_{1}$ and $\mathcal{D}_{2}$. Then $\mathcal{D}_{1}$ is used to select the values of the 7 hyperparameters of the LSTM models, and $\mathcal{D}_{2}$ is used to estimate the conditional densities using the LSTM models specified by the selected hyperparameters. In $\mathcal{D}_{1}$, we select the hyperparameters as described in the previous section.
In $\mathcal{D}_{2}$, the LSTM models are fit with the hyperparameter values obtained on $\mathcal{D}_{1}$ and used to estimate the conditional densities of covariates and outcome in the NICE g-formula.

In contrast to the typical use of sample splitting in the literature -- where nuisance functions are fit on one subsample and the causal parameter is estimated on an independent subsample -- we use sample splitting to separate model selection from causal estimation. Specifically, hyperparameter tuning of the LSTM models is conducted in the subset $\mathcal{D}_{1}$ and, after hyperparameters are selected, the nuisance models are trained and the NICE g-formula estimand is evaluated on the independent subset $\mathcal{D}_{2}$. 
This approach prevents observations used for causal estimation from also being used for hyperparameter selection, thereby limiting data reuse associated with model selection~\cite{cawley2010overfitting}. Moreover, by treating the selected hyperparameters as fixed during estimation, this strategy substantially reduces the computational burden of bootstrap-based inference, since hyperparameter tuning can be performed once and reused across bootstrap replicates rather than repeated for each resampled dataset.

Subsequently, same as for the parametric g-formula, the sum over all covariate histories is approximated via Monte Carlo simulation, and the standard error of the estimates is estimated via nonparametric bootstrapping.


We then apply cross-fitting with $\mathcal{D}_{2}$ to select the values of the hyperparameters of the LSTM models, and $\mathcal{D}_{1}$ used to estimate the conditional densities using the LSTMs and compute the g-formula estimate. The final g-formula risk estimate $\hat{p}$ is the average of the risk estimates $\hat{p}_1$ and $\hat{p}_2$ from $\mathcal{D}_{1}$ and $\mathcal{D}_{2}$, respectively. The standard error estimate $\hat{s}$ is $\sqrt{\frac{(\hat{s}_1^2+\hat{s}_2^2)}4}$, where $\hat{s}_1$ and $\hat{s}_2$ are the bootstrap estimates of the standard error in $\mathcal{D}_{1}$ and $\mathcal{D}_{2}$, respectively. The 95\% confidence interval for the risk estimate is constructed via the normal approximation
$ \hat{p} \pm z_{0.975} \cdot \hat{s}$ where $z_{0.975}$ denotes the $0.975$ quantile of the standard normal distribution.
The overall procedure is outlined in Algorithm \ref{alg:sample-splitting}.

\section{Simulation study}
\subsection{Data generation}
\label{subsec:simulation}

We simulated a cohort of people with HIV who started antiretroviral treatment for the first time. We used this simulated dataset to mimic a recent study that estimated the effect of integrase inhibitors (INSTIs; a class of antiretrovirals) on the risk of cardiovascular events \cite{rein2023integrase} in the HIV-CAUSAL and ART-CC Collaborations, two consortia of observational cohorts of people living with HIV from Europe and North America \cite{Lodi}. 
We generated datasets with a time-varying treatment $A_k$ (1: use of an antiretroviral regimen that contains an INSTI drug, 0: otherwise), a time-varying outcome $Y_k$ (1: cardiovascular event, 0: otherwise), and the baseline confounders $L_0$ and time-varying confounders $L_k$:


\begin{itemize}
\def\labelitemi{$\bullet$}

    \item Baseline sex (1=male, 0=female),
    \item Baseline age (continuous),
    \item Baseline smoking (0=never smoked, 1=currently smoking, 2=ex-smoker),
    \item Time-varying CD4 cell count (continuous),
    \item Time-varying HIV RNA viral load (continuous),
    \item Time-varying BMI (1=overweight or obese, 0=otherwise).

\end{itemize}

The data were generated to emulate the structure and characteristics of real-world longitudinal HIV data, as described in Appendix~\ref{appendix:datagen}. Specifically, we considered two data-generating scenarios: a simpler scenario, in which the covariates and outcome depend only on the one-step lagged history, and a more complex scenario, in which the covariates and outcome depend on the entire observed history up to that time point.
For both scenarios, we simulated data for each individual at months $k = 0, 1, 2, \dots, K$, where $k = 0$ is baseline and $K = 9$. For each individual and month, we simulated a time-varying indicator for treatment $A_k$ during month $k$, a vector of baseline confounders at $k=0$, a vector of time-varying confounders at each month $k$ and a time-varying indicator of the outcome $Y_k$ during month $k$. For simplicity, in the simulated data, no individual was lost to follow-up or died from other causes. 

\subsection{Analysis methods}
\label{subsec:analysis}

For both the simple and complex HIV data scenarios, we used both parametric and LSTM-based NICE g-formula estimators to estimate the 10-month risk under four interventions:

\begin{itemize}
\def\labelitemi{$\bullet$}
    \item Natural course: no intervention on each individual's observed treatment over time. 
    \item Always treat: set treatment to 1 for each individual at all time points.
    \item Never treat: set treatment to 0 for each individual at all time points.
    \item Dynamic intervention: for each individual, initiate treatment when the current RNA level exceeds the threshold of 200 copies/ml and maintain treatment at 1 at all subsequent time points.
\end{itemize}

To approximate the true value of the risk under each intervention, we simulated a large reference dataset of 10 million individuals under the same data-generating mechanism.

For evaluation purposes, we generated simulated datasets under two different sample sizes: 1,000 datasets with 1,000 individuals each and 1,000 datasets with 10,000 individuals each. Because estimators are sensitive to the scale of the input variables, we applied variable transformations to improve the stability of the estimator. See Appendix~\ref{app:implementation} for further details on data preparation and preprocessing used in our analysis.

In each dataset, we used LSTM-based and parametric estimators with 10,000 Monte Carlo simulations to estimate the risk under each intervention and construct 95\% confidence interval around the risk estimate using a normal approximation, with standard errors estimated from 300 bootstrap samples.

We estimated the bias of our estimators as the average difference between the estimated risk and the true risk across the 1,000 datasets. We also computed the standard error of the estimated bias and its 95\% confidence interval using a normal approximation. Variance was computed as the empirical sampling variance of the estimator across the 1,000 simulation datasets.
We estimated coverage as the proportion of the 1,000 datasets in which the 95\% confidence interval around the risk estimate contained the true risk.
Moreover, for each intervention, we also estimate the risk difference and risk ratio relative to the natural course, along with their standard errors, 95\% confidence intervals, and empirical coverage probabilities.

We applied the following estimators to each dataset.

\subsubsection{Parametric estimators}
We considered two parametric estimators that differ in how closely the models track the parameterization of the measured variables in the models used to generate the data.

\begin{itemize}
\def\labelitemi{$\bullet$}
\item More misspecified estimator: For the simpler longitudinal data, the parametric models include the lagged value of each time-varying covariate, the baseline covariates and a variable for time. They do not include the nonlinear terms and product ("interaction") terms in the covariate and outcome models used to generate the data. Similarly, for the more complex data, the parametric models include the lagged value of each time-varying covariate, baseline covariates and a time variable. They do not include the nonlinear terms and interaction terms in the covariate and outcome models used to generate the data.
\item Less misspecified estimator: For the simpler longitudinal data, besides the lagged value of each time-varying covariate and baseline covariates, the parametric models also include the nonlinear terms and interaction terms, and a time-dependent interaction term is included to allow the coefficients for the lagged values of CD4 and RNA to change after six months as in the data generation. 
For the more complex longitudinal data, the parametric models include the nonlinear and interaction terms, as well as lagged values and lagged cumulative averages of the time-varying covariates over all prior time points, together with baseline covariates. Moreover, time-dependent interaction terms are included to allow the coefficients for the lagged values and lagged cumulative averages of CD4 and RNA to change after six months as in the data generation.

\end{itemize}

See Appendix \ref{supp:model_spec} for the full parameterization of the parametric g-formula estimators.

\subsubsection{LSTM-based estimators}
\def\labelitemi{$\bullet$}
We consider two LSTM-based NICE estimators:
\begin{itemize}
\item No sample splitting: The full dataset is used for both hyperparameter tuning and g-formula estimation.
\item Sample splitting and cross-fitting: A random half of the dataset is used for hyperparameter tuning, and the other half for estimation. In addition, cross-fitting is applied, in which the two data halves are alternated for hyperparameter tuning and g-formula estimation.

\end{itemize}

\section{Results}
\label{sec:results}

Table \ref{tab:simple_risk_all_interventions} and Table \ref{tab:complex_risk_all_interventions} present the estimated risks from the parametric and LSTM-based g-formula estimators, along with their bias, variance, and coverage under the natural course and each of the interventions. Table \ref{tab:simple_risk_all_interventions} presents the results from simulations using the simpler data generation mechanism, and Table \ref{tab:complex_risk_all_interventions} using the more complex data generation mechanism. 

\begin{table}[!htb]
\centering

\caption{\textbf{Simple longitudinal HIV data.}
Performance of parametric and LSTM-based NICE g-formula estimators
under four interventions.}
\label{tab:simple_risk_all_interventions}

\small
\setlength{\tabcolsep}{4pt}
\renewcommand{\arraystretch}{1.12}

\resizebox{\textwidth}{!}{%
\begin{tabular}{clccc>{\centering\arraybackslash}p{2.4cm}cc}
\toprule
Sample size
& Estimator
& Risk ($\%$)
& \multicolumn{3}{c}{Bias ($\%$)}
& Variance ($\%^{2}$) 
& Coverage ($\%$) \\
\cmidrule(lr){4-6}
&
&
& Estimate
& Std.\ error
& 95\% CI
&
& \\
\hline

\multicolumn{8}{l}{
\textbf{Natural course}
\hspace{1em}(true risk = 14.52\%)
} \\
\addlinespace[2pt]

\multirow{4}{*}{1,000}
& LSTM with sample splitting
& 14.47 & -0.06 & 0.04 & (-0.14, 0.03) & 1.93 & 93.9 \\
& LSTM without sample splitting
& 14.22 & -0.30 & 0.04 & (-0.38, -0.22) & 1.66 & 95.8 \\
& Parametric with larger misspecification
& 15.33 & 0.81 & 0.04 & (0.73, 0.88) & 1.34 & 90.5 \\
& Parametric with smaller misspecification
& 15.00 & 0.47 & 0.04 & (0.40, 0.54) & 1.28 & 93.5 \\
\addlinespace[3pt]

\multirow{4}{*}{10,000}
& LSTM with sample splitting
& 14.19 & -0.33 & 0.02 & (-0.36, -0.30) & 0.23 & 96.9 \\
& LSTM without sample splitting
& 14.19 & -0.34 & 0.02 & (-0.37, -0.30) & 0.28 & 97.7 \\
& Parametric with larger misspecification
& 15.32 & 0.80 & 0.01 & (0.78, 0.82) & 0.12 & 44.0 \\
& Parametric with smaller misspecification
& 15.00 & 0.47 & 0.01 & (0.45, 0.50) & 0.12 & 75.1 \\

\hline

\multicolumn{8}{l}{
\textbf{Always treat}
\hspace{1em}(true risk = 11.75\%)
} \\
\addlinespace[2pt]

\multirow{4}{*}{1,000}
& LSTM with sample splitting
& 11.72 & -0.03 & 0.05 & (-0.12, 0.07) & 2.25 & 92.5 \\
& LSTM without sample splitting
& 11.36 & -0.39 & 0.04 & (-0.47, -0.30) & 1.89 & 93.7 \\
& Parametric with larger misspecification
& 12.22 & 0.47 & 0.04 & (0.40, 0.55) & 1.59 & 94.0 \\
& Parametric with smaller misspecification
& 12.06 & 0.31 & 0.04 & (0.24, 0.39) & 1.54 & 94.1 \\
\addlinespace[3pt]

\multirow{4}{*}{10,000}
& LSTM with sample splitting
& 11.53 & -0.22 & 0.02 & (-0.25, -0.19) & 0.25 & 96.1 \\
& LSTM without sample splitting
& 11.54 & -0.21 & 0.02 & (-0.24, -0.17) & 0.28 & 97.8 \\
& Parametric with larger misspecification
& 12.19 & 0.45 & 0.01 & (0.42, 0.47) & 0.15 & 82.3 \\
& Parametric with smaller misspecification
& 12.05 & 0.30 & 0.01 & (0.27, 0.32) & 0.14 & 90.3 \\

\hline

\multicolumn{8}{l}{
\textbf{Never treat}
\hspace{1em}(true risk = 20.88\%)
} \\
\addlinespace[2pt]

\multirow{4}{*}{1,000}
& LSTM with sample splitting
& 20.73 & -0.16 & 0.09 & (-0.34, 0.03) & 8.50 & 92.5 \\
& LSTM without sample splitting
& 20.74 & -0.14 & 0.09 & (-0.33, 0.04) & 8.98 & 93.5 \\
& Parametric with larger misspecification
& 22.53 & 1.64 & 0.08 & (1.48, 1.80) & 6.80 & 91.3 \\
& Parametric with smaller misspecification
& 21.83 & 0.95 & 0.08 & (0.79, 1.10) & 6.32 & 93.5 \\
\addlinespace[3pt]

\multirow{4}{*}{10,000}
& LSTM with sample splitting
& 20.25 & -0.63 & 0.03 & (-0.70, -0.57) & 1.02 & 94.5 \\
& LSTM without sample splitting
& 20.28 & -0.60 & 0.03 & (-0.67, -0.53) & 1.18 & 94.5 \\
& Parametric with larger misspecification
& 22.56 & 1.68 & 0.03 & (1.63, 1.73) & 0.67 & 48.0 \\
& Parametric with smaller misspecification
& 21.87 & 0.99 & 0.02 & (0.94, 1.04) & 0.62 & 77.4 \\

\hline

\multicolumn{8}{l}{
\textbf{Dynamic intervention}
\hspace{1em}(true risk = 16.53\%)
} \\
\addlinespace[2pt]

\multirow{4}{*}{1,000}
& LSTM with sample splitting
& 16.72 & 0.19 & 0.06 & (0.08, 0.30) & 3.15 & 95.4 \\
& LSTM without sample splitting
& 16.51 & -0.03 & 0.06 & (-0.13, 0.08) & 3.06 & 97.1 \\
& Parametric with larger misspecification
& 17.67 & 1.14 & 0.05 & (1.05, 1.23) & 2.23 & 90.2 \\
& Parametric with smaller misspecification
& 17.23 & 0.70 & 0.05 & (0.61, 0.79) & 2.09 & 93.1 \\
\addlinespace[3pt]

\multirow{4}{*}{10,000}
& LSTM with sample splitting
& 16.20 & -0.33 & 0.02 & (-0.37, -0.29) & 0.39 & 97.4 \\
& LSTM without sample splitting
& 16.22 & -0.31 & 0.02 & (-0.36, -0.27) & 0.48 & 96.9 \\
& Parametric with larger misspecification
& 17.66 & 1.12 & 0.01 & (1.10, 1.15) & 0.22 & 36.9 \\
& Parametric with smaller misspecification
& 17.23 & 0.70 & 0.01 & (0.67, 0.73) & 0.20 & 67.6 \\

\bottomrule
\end{tabular}
}
\end{table}

\begin{table}[!htb]
\centering

\caption{\textbf{Complex longitudinal HIV data.}
Evaluation of parametric and LSTM-based NICE g-formula estimators
under four interventions.}
\label{tab:complex_risk_all_interventions}

\small
\setlength{\tabcolsep}{4pt}
\renewcommand{\arraystretch}{1.12}

\resizebox{\textwidth}{!}{%
\begin{tabular}{clcccccc}
\toprule
Sample size
& Estimator
& Risk ($\%$)
& \multicolumn{3}{c}{Bias ($\%$)}
& Variance ($\%^2$)
& Coverage ($\%$) \\
\cmidrule(lr){4-6}
&
&
& Estimate
& Std.\ error
& 95\% CI
&
& \\
\hline

\multicolumn{8}{l}{
\textbf{Natural course}
\hspace{1em}(true risk = 16.80\%)
} \\
\addlinespace[2pt]

\multirow{4}{*}{1,000}
& LSTM with sample splitting
& 17.30 & 0.50 & 0.05 & (0.40, 0.59) & 2.22 & 92.0 \\
& LSTM without sample splitting
& 16.75 & -0.05 & 0.05 & (-0.15, 0.04) & 2.35 & 94.6 \\
& Parametric with larger misspecification
& 10.54 & -6.26 & 0.03 & (-6.32, -6.20) & 1.05 & 0.0 \\
& Parametric with smaller misspecification
& 14.61 & -2.19 & 0.04 & (-2.27, -2.12) & 1.47 & 54.5 \\
\addlinespace[3pt]

\multirow{4}{*}{10,000}
& LSTM with sample splitting
& 16.56 & -0.24 & 0.02 & (-0.28, -0.20) & 0.39 & 94.4 \\
& LSTM without sample splitting
& 16.64 & -0.16 & 0.02 & (-0.20, -0.13) & 0.39 & 96.1 \\
& Parametric with larger misspecification
& 10.61 & -6.19 & 0.01 & (-6.21, -6.17) & 0.10 & 0.0 \\
& Parametric with smaller misspecification
& 14.72 & -2.08 & 0.01 & (-2.11, -2.06) & 0.14 & 0.0 \\

\hline

\multicolumn{8}{l}{
\textbf{Always treat}
\hspace{1em}(true risk = 12.81\%)
} \\
\addlinespace[2pt]

\multirow{4}{*}{1,000}
& LSTM with sample splitting
& 12.51 & -0.30 & 0.05 & (-0.39, -0.20) & 2.42 & 93.7 \\
& LSTM without sample splitting
& 12.44 & -0.37 & 0.05 & (-0.47, -0.27) & 2.43 & 92.9 \\
& Parametric with larger misspecification
& 9.99 & -2.82 & 0.04 & (-2.89, -2.74) & 1.46 & 37.4 \\
& Parametric with smaller misspecification
& 12.63 & -0.17 & 0.04 & (-0.26, -0.09) & 1.91 & 94.4 \\
\addlinespace[3pt]

\multirow{4}{*}{10,000}
& LSTM with sample splitting
& 12.53 & -0.28 & 0.02 & (-0.31, -0.24) & 0.38 & 93.9 \\
& LSTM without sample splitting
& 12.64 & -0.17 & 0.02 & (-0.21, -0.13) & 0.42 & 96.2 \\
& Parametric with larger misspecification
& 10.06 & -2.75 & 0.01 & (-2.77, -2.73) & 0.14 & 0.0 \\
& Parametric with smaller misspecification
& 12.72 & -0.09 & 0.01 & (-0.11, -0.06) & 0.19 & 94.9 \\

\hline

\multicolumn{8}{l}{
\textbf{Never treat}
\hspace{1em}(true risk = 21.29\%)
} \\
\addlinespace[2pt]

\multirow{4}{*}{1,000}
& LSTM with sample splitting
& 22.73 & 1.44 & 0.06 & (1.31, 1.57) & 4.20 & 86.0 \\
& LSTM without sample splitting
& 21.61 & 0.32 & 0.07 & (0.19, 0.45) & 4.44 & 93.1 \\
& Parametric with larger misspecification
& 11.17 & -10.12 & 0.05 & (-10.21, -10.03) & 2.04 & 0.0 \\
& Parametric with smaller misspecification
& 16.90 & -4.39 & 0.05 & (-4.49, -4.29) & 2.54 & 21.0 \\
\addlinespace[3pt]

\multirow{4}{*}{10,000}
& LSTM with sample splitting
& 21.08 & -0.21 & 0.03 & (-0.26, -0.16) & 0.64 & 94.3 \\
& LSTM without sample splitting
& 21.14 & -0.14 & 0.03 & (-0.19, -0.09) & 0.65 & 96.7 \\
& Parametric with larger misspecification
& 11.23 & -10.05 & 0.01 & (-10.08, -10.03) & 0.20 & 0.0 \\
& Parametric with smaller misspecification
& 17.03 & -4.25 & 0.02 & (-4.28, -4.22) & 0.24 & 0.0 \\

\hline

\multicolumn{8}{l}{
\textbf{Dynamic intervention}
\hspace{1em}(true risk = 13.68\%)
} \\
\addlinespace[2pt]

\multirow{4}{*}{1,000}
& LSTM with sample splitting
& 13.85 & 0.17 & 0.05 & (0.08, 0.26) & 2.21 & 93.4 \\
& LSTM without sample splitting
& 13.51 & -0.17 & 0.05 & (-0.26, -0.08) & 2.15 & 95.0 \\
& Parametric with larger misspecification
& 10.09 & -3.59 & 0.03 & (-3.65, -3.53) & 0.97 & 6.8 \\
& Parametric with smaller misspecification
& 13.12 & -0.57 & 0.04 & (-0.64, -0.49) & 1.37 & 91.3 \\
\addlinespace[3pt]

\multirow{4}{*}{10,000}
& LSTM with sample splitting
& 13.48 & -0.21 & 0.02 & (-0.24, -0.17) & 0.36 & 95.2 \\
& LSTM without sample splitting
& 13.58 & -0.10 & 0.02 & (-0.14, -0.06) & 0.38 & 97.1 \\
& Parametric with larger misspecification
& 10.16 & -3.52 & 0.01 & (-3.54, -3.50) & 0.09 & 0.0 \\
& Parametric with smaller misspecification
& 13.21 & -0.47 & 0.01 & (-0.49, -0.45) & 0.13 & 74.6 \\

\bottomrule
\end{tabular}%
}
\end{table}


Across interventions and data generation mechanisms, the LSTM-based estimators usually had lower absolute bias than the parametric estimators, but their variances were larger, as expected.
The LSTM-based estimator had near nominal coverage, regardless of whether sample splitting was used, across all scenarios when the sample size was $n=10{,}000$. In contrast, the parametric estimators had poor or very poor coverage for complex data generation mechanisms. 

\begin{figure}[!htb]
    \centering
    \begin{subfigure}{\textwidth}
        \centering
        \includegraphics[width=\textwidth]{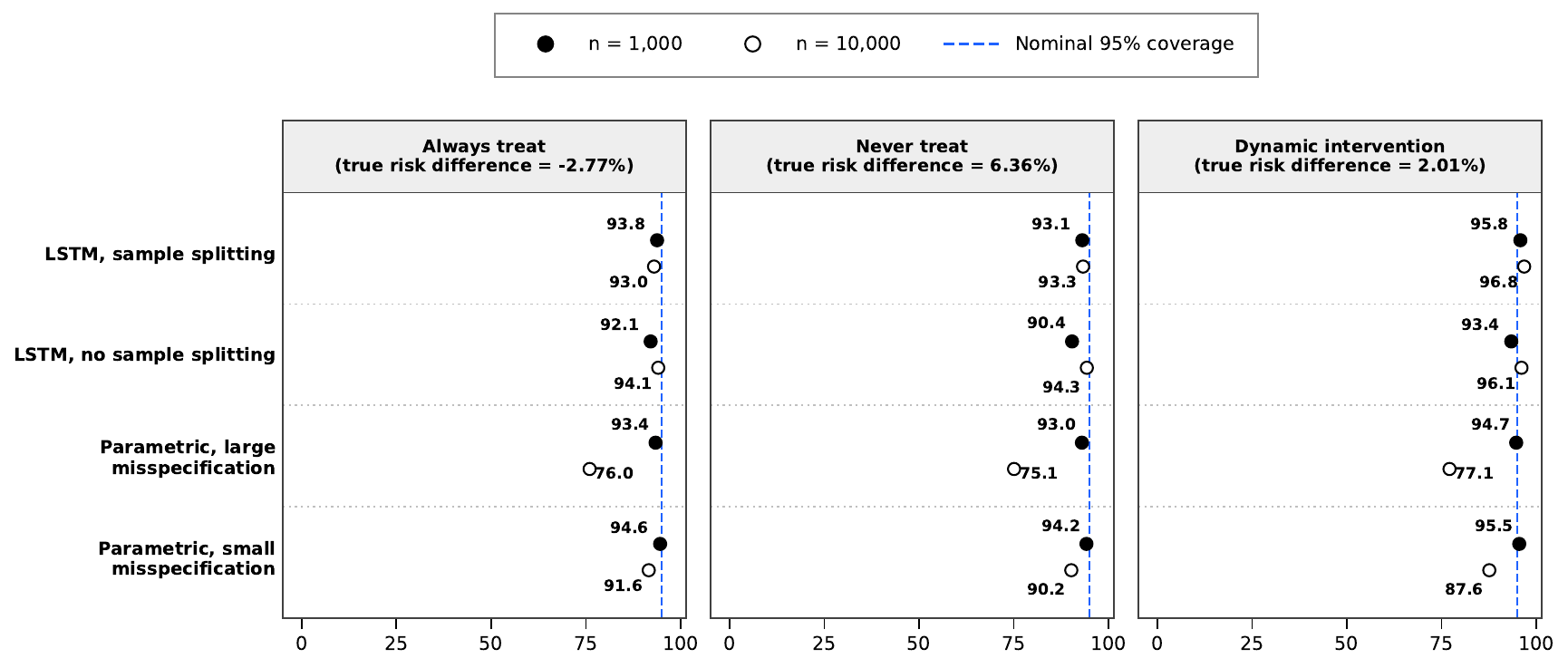}
        \caption{Risk difference}
        \label{fig:simple_rd_coverage}
    \end{subfigure}

    \vspace{0.5em}

    \begin{subfigure}{\textwidth}
        \centering
        \includegraphics[width=\textwidth]{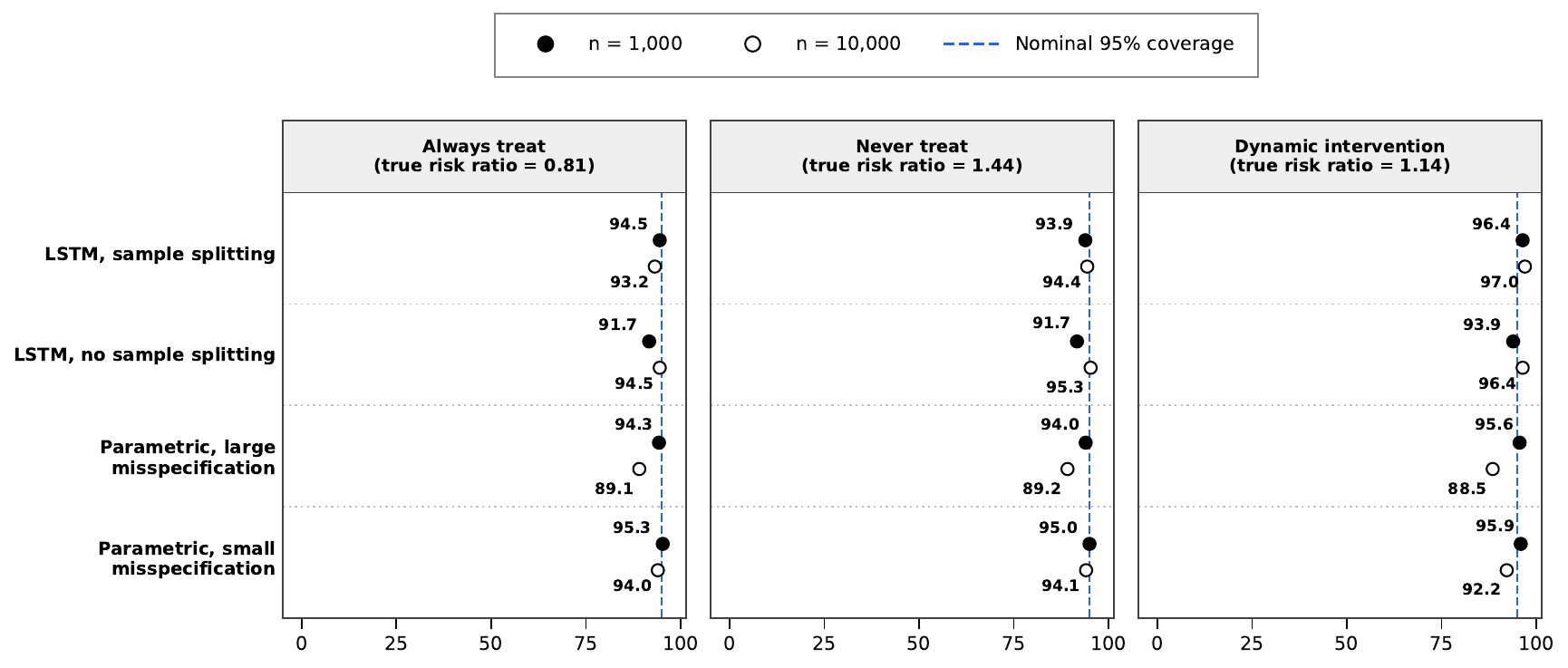}
        \caption{Risk ratio}
        \label{fig:simple_rr_coverage}
    \end{subfigure}

    \caption{Empirical coverage of nominal 95\% confidence intervals for risk differences and risk ratios in the simple longitudinal HIV data.}
    \label{fig:simple_rd_rr_coverage}
\end{figure}

\begin{figure}[!htb]
    \centering
    \begin{subfigure}{\textwidth}
        \centering
        \includegraphics[width=\textwidth]{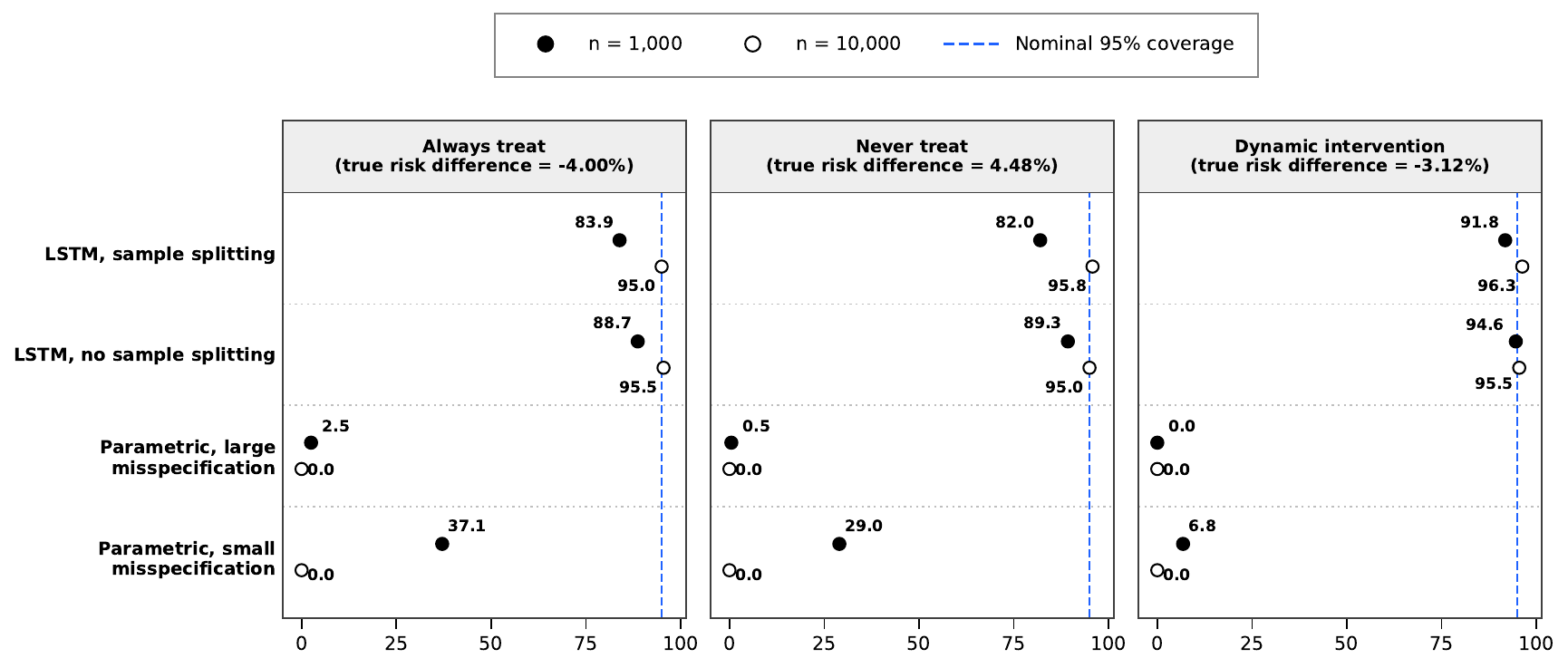}
        \caption{Risk difference}
        \label{fig:complex_rd_coverage}
    \end{subfigure}

    \vspace{0.5em}

    \begin{subfigure}{\textwidth}
        \centering
        \includegraphics[width=\textwidth]{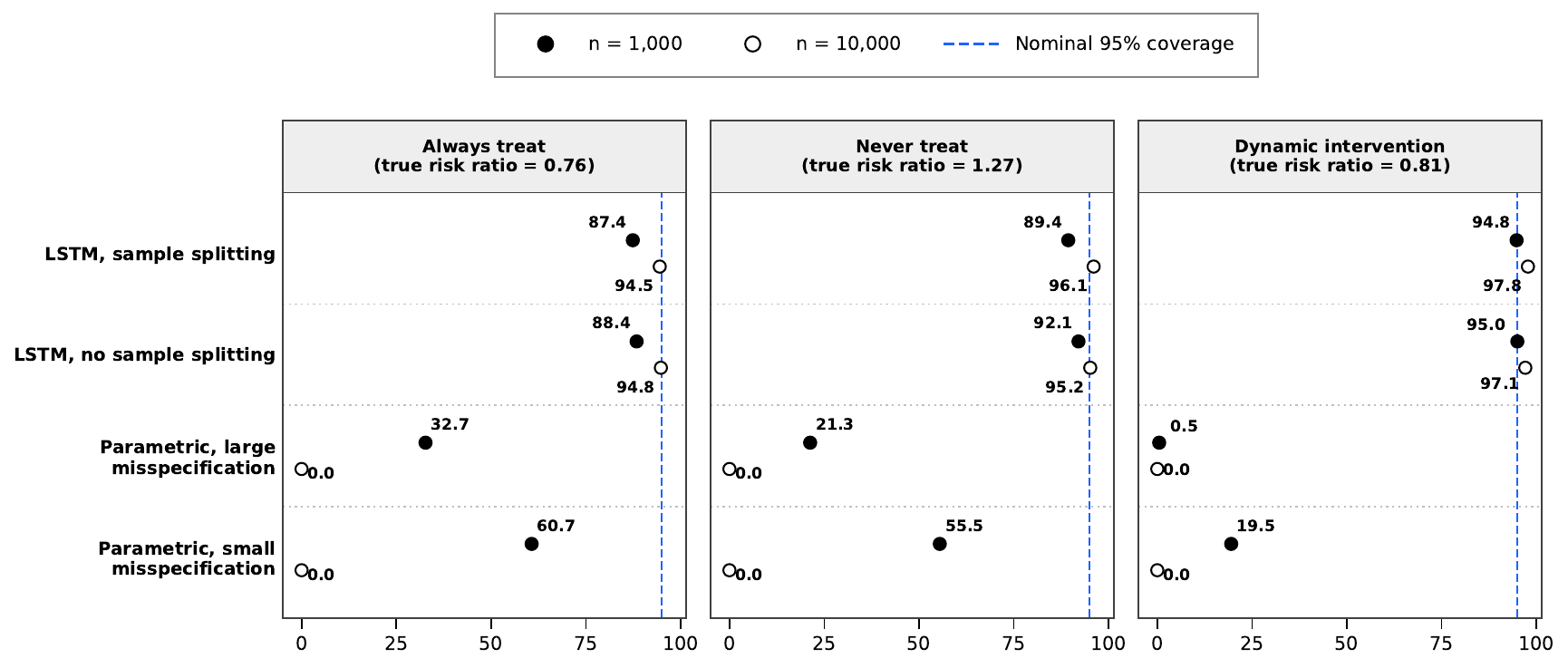}
        \caption{Risk ratio}
        \label{fig:complex_rr_coverage}
    \end{subfigure}

    \caption{Empirical coverage of nominal 95\% confidence intervals for risk differences and risk ratios in the complex longitudinal HIV data.}
    \label{fig:complex_rd_rr_coverage}
\end{figure}

Figure~\ref{fig:simple_rd_rr_coverage} and Figure~\ref{fig:complex_rd_rr_coverage} show the empirical coverage probabilities for risk differences and risk ratios, using the natural course as the reference, for the simple and complex longitudinal data, respectively. The LSTM-based estimators achieved coverage close to the nominal 95\% level for risk differences and risk ratios in all settings when the sample size was 10,000, but not always for the sample size of 1,000. In contrast, the parametric g-formula estimators generally had lower coverage, particularly under the complex data generation mechanism. (See Appendix~\ref{supp:risk_ratio_difference} for more detailed quantitative results.)

The time required to run the parametric g-formula estimator is in the order of minutes per replicate. In contrast, the LSTM-based estimator incurs substantially higher computational cost, with per-replicate runtimes in the order of hours, reflecting a trade-off between computational efficiency and modeling flexibility. To mitigate this cost, we employed parallelization to train multiple models concurrently on available GPU resources. (See Appendix \ref{app:implementation} for more details on the implementation.)





\section{Discussion}
\label{sec:discussion}

Our simulations illustrate that, as data complexity and sample size increase, LSTM-based estimators of the NICE g-formula become substantially less biased than parametric estimators constructed with partial knowledge of the data generation mechanism. This finding was expected because the LSTM-based g-formula estimator does not rely on prespecified functional forms and instead uses neural networks to flexibly capture dependencies across variables and time. Conversely, in a bias-variance tradeoff, the variance of the LSTM-based estimators is larger than that of parametric estimators. 

Importantly, the LSTM-based estimators achieved near-nominal coverage for the 10,000 sample size. This finding is not predicted by existing theory and thus raises interesting theoretical questions for future pursuit. Specifically, as noted in Section \ref{subsec: sample_splitting}, recent literature has provided rigorous analyses of plug-in principles and their validity in specific high-dimensional instances. The results here lay open newer examples for which such rigorous high-dimensional analyses might be valuable beyond the linear models typically considered in literature \citep{yadlowsky2022explaining,mcgrath2025optimal}.

Our study also explores a specific way of tuning nuisance functions through a cross-validation approach. Recent work in high-dimensional problems \citep{mcgrath2025optimal,jiang2025new} argue that choosing tuning parameters for nuisance learning by directly optimizing the asymptotic variance of the causal effect estimators can have substantive gains over the standard choice of tuning selection from a prediction perspective. Exploring this line of work in the context of NICE g-formula-based estimation remains an open direction for future research.



\vspace{3em}
\noindent\textbf{Acknowledgements:}
The computations in this manuscript were run in part on the FAS Research Computing (FASRC) Cannon cluster, supported by the FAS Division of Science Research Computing Group at Harvard University.

\par\vspace{1.5em}

\noindent\textbf{Conflict of interest:}
The authors have no conflicts of interest to declare.
\par\vspace{1.5em}

\noindent\textbf{Research funding:} This research was supported by U.S. National Institutes of Health grant R37 AI102634, U.S. Office of Naval Research MURI grant N000142412687, and an ADIA Lab Health Sciences Grant.
\par\vspace{1.5em}

\noindent\textbf{Author contributions:}
All authors accepted the responsibility for the content of the manuscript and consented to its submission, reviewed all the results, and approved the final version of the manuscript. 
\par\vspace{1.5em}

\noindent\textbf{Data availability:} This study used simulated data. The data-generating function and simulation parameters are provided in Appendix~\ref{appendix:datagen}.
\par\vspace{1.5em}

\appendix
\section*{Appendix}

\section{Identifying assumptions}
\label{appendix:identify}

Let $g = (g_0, ..., g_K)$ be a treatment strategy on $A_k$ that depends on the measured covariate and treatment history, characterized by the intervention density $f^{\text{int}}(a_k \mid \bar{l}_{k}, \bar{a}^g_{k-1})$. Let $Y^g_k$ and $L^g_k$
denote the counterfactual outcome and covariates, respectively, at time $k$ had an individual followed a particular treatment strategy $g$. We consider the cumulative incidence (risk) over a specified follow-up period from time 0 to $K$ under a sustained treatment strategy in the survival setting. The causal estimand can be identified under the following assumptions \cite{hernan2020causal}:

\begin{itemize}

\item  Exchangeability. For each $k \in \{0, 1, ..., K\}$, 

\[
(Y^{g}_{k+1}, \ldots, Y^{g}_{K}) \perp\!\!\!\perp A_k \,|\, \bar{L}_k = \bar{l}_k, \bar{A}_{k-1} = \bar{a}^{g}_{k-1}, Y_k = 0
\]

The exchangeability assumption ensures that the potential outcome $Y^g_k$ under treatment $A^g_k$ at each interval k is independent of the actual treatment $A_k$ conditional on the confounder and treatment history $(\bar{L_k}, \bar{A_k})$.

\item  Positivity. 
\begin{align*}
& f_{\bar{L}_k, \bar{A}_{k-1}, Y_k}(\bar{l}_k, \bar{a}^g_{k-1}, 0) > 0 \\
\Rightarrow & \quad f^{obs}(a^g_k \mid \bar{l}_k, \bar{a}^g_{k-1}, 0) = f_{A_k\mid \bar{L}_k, \bar{A}_{k-1}, Y_k}(a^g_k \mid \bar{l}_k, \bar{a}^g_{k-1}, 0) > 0
\end{align*}

The positivity assumption ensures that there exists a non-zero probability that the individual being assigned to the treatment following the strategy $g$ given the confounder history, up to each time $k$.

\item  Consistency. If $\bar{A_k} = \bar{a}^g_k$, then $Y_{k+1} = Y^g_{k+1}$ and $\bar{L}_k = \bar{L}^g_k$.

\end{itemize}

If these assumptions hold, for all $g \in \mathcal{G}$ where $\mathcal{G}$ is a set of deterministic treatment strategies, the risk by the end of follow up had all individuals been assigned treatment according to $f^{\text{int}}(a_k \mid \bar{l}_{k}, \bar{a}_{k-1})$ is equivalent to the non-iterative conditional expectation g-formula characterized by the observed densities of the outcome, confounders and the intervention density of the treatment:

\begin{align*}
& \sum_{\bar{a}_K} \sum_{\bar{l}_K} \sum_{k=0}^{K} 
\Pr\left(Y_{k+1} = 1 \mid \bar{L}_k = \bar{l}_k, \bar{A}_k = \bar{a}_k, \bar{Y}_k = 0\right) \times \\
& \quad \prod_{j=0}^{k}
\Big\{ 
f\left(l_j \mid \bar{l}_{j-1}, \bar{a}_{j-1}\right) 
f^{\text{int}}\left(a_j \mid \bar{l}_j, \bar{a}_{j-1}\right) 
\times \\
& \quad \Pr\left(Y_j = 0 \mid \bar{L}_{j-1} = \bar{l}_{j-1}, \bar{A}_{j-1} = \bar{a}_{j-1},\bar{Y}_{j-1} = 0\right)
\Big\}
\end{align*}

\section{Implementation details}
\label{app:implementation}

\textbf{Dataset preparation.} Two datasets are constructed using the standardized training and validation data for training the covariate and outcome models separately:
(1) Covariate dataset: For each individual $i$, the input sequence consists of covariate values from time 
$k=0$ to $k=K_{i}-1$, and the output sequence contains covariate values from time 
$k=1$ to $k=K_{i}$.
(2) Outcome dataset: For each individual $i$, the input sequence includes the covariate history observed from $k=0$ through the end of follow-up at $k=K_{i}$, and the corresponding output sequence comprises the binary event outcomes observed from $k=1$ to $k=K_{i}+1$. Here, \(K_i\) denotes the individual-specific final follow-up time for individual \(i\). If individual \(i\) remains alive and uncensored through the end of administrative follow-up, then \(K_i=K\).

\textbf{Data preprocessing.} We applied transformations to the original input dataset. Binary variables (sex, BMI status, INSTI use, and the outcome) were retained in their form and the categorical variable smoking was transformed into indicator variables using one-hot-encoding. Continuous variables (CD4 cell count, HIV-RNA viral load, age) were standardized to have zero mean and unit variance. For each variable, the mean and standard deviation were calculated using data from all individuals across all time points in the training set. 
These transformations were applied to both the training and validation sets.
For individuals who did not survive until the end of follow-up, covariate values after death were padded with a predefined placeholder. This placeholder was used to construct a mask during model training, allowing the model to exclude those positions from loss computation.

\textbf{Model training details.} During training of the covariate model, the binary covariates are optimized using binary cross-entropy loss, and continuous covariates using mean squared error. The total training loss is computed as the sum of the individual covariate losses. A binary mask is created to identify valid (non-padded) positions in the target output tensor of the covariates. Padding values are excluded from the loss computation. An early stopping mechanism is applied to halt training if the validation loss does not improve after 50 epochs. When training the outcome model, the outcome is optimized using binary cross-entropy loss. A binary mask is created to identify valid (non-padded) positions in the target output tensor of the outcome. Both models are optimized using the AdamW optimizer. The maximum number of training epochs is set to 10,000, and the patience epoch for early stopping is set to 50.

For the LSTM model estimating the conditional distribution of covariates, the network outputs the predicted conditional mean for each continuous covariate. 
These covariates are subsequently generated by adding residuals sampled from the empirical residual distribution to the corresponding LSTM outputs.
For the LSTM model estimating the conditional distribution of the outcome, the network outputs the predicted conditional probability of the outcome at each time point. These probabilities are then combined to compute the final estimated risk.

\textbf{Computational implementation.} We generate 1,000 random seeds, each of the 1,000 simulated datasets is generated using a distinct random seed. In the 1,000 simulation replicates conducted for the LSTM-based estimator, each replicate is executed on a single GPU. Within each replicate, hyperparameter tuning consisted of 50 trials, which were executed in parallel on the assigned GPU in batches of 10 or 15 trials, depending on the GPU memory capacity. For bootstrap inference, we generated 300 bootstrap resamples, with model fitting parallelized by training 10 or 15 bootstrap models concurrently on the same GPU. When training the model on each bootstrap resample, the neural network was initialized with the weights from the model trained on the original observed dataset.

\textbf{Computer resources.} Analyses were executed on three GPU clusters equipped with heterogeneous NVIDIA GPUs. All simulations were run in computing environments using Python 3.8–3.13, CUDA Toolkit 12.1–12.6, and PyTorch 2.4.1–2.7.0. The GPU models included NVIDIA A40, A100, RTX A6000, H100, H200, Tesla V100, Tesla V100S, RTX 2080 Ti, RTX 3060, RTX 3090, RTX 4060 Ti, RTX 4070S, RTX 4080, RTX 4080S, RTX 4090, RTX 5060 Ti, and RTX 5080.

\section{Loss calculation}
\label{app:loss}

In the following, we describe the details of the training and validation loss calculation for both the covariate and outcome models.

When training the covariate model, the loss function is defined as:

\[
loss_{L_i} \;=\; 
\frac{
\displaystyle \sum_{b=1}^{B} \sum_{t=1}^{T} 
m_{b,t,i} \; \ell_i\!\left(\hat{l}_{b,t,i},\, l_{b,t,i}\right)
}{
\displaystyle \sum_{b=1}^{B} \sum_{t=1}^{T} m_{b,t,i}
},
\]

\[
\text{covariates\_loss}_{train}
\;=\;
\sum_{i=1}^{C} loss_{L_i}.
\]

where $B$ denotes the number of individuals in each training batch, $T$ denotes the maximum sequence length, $C$ denotes the total number of all covariates. $m_{b, t, i}$ is a mask indicating if the $i$-th covariate has missing values at time point $t$ for the $b$-th individual in the training batch. $\ell(\cdot)$ denotes the covariate-specific loss function, where
$\ell(\cdot)$ is the mean squared error (MSE) loss for continuous covariates
and the binary cross-entropy (BCE) loss for binary covariates.

On the validation set, the total loss function for all covariates is defined as:

\[
loss_{L_i}
=
\frac{
\displaystyle
\sum_{n=1}^{N}
\sum_{t=1}^{T}
m_{n,t,i}\,
\ell_i\!\left(\hat{l}_{n,t,i},\, l_{n,t,i}\right)
}{
\displaystyle
\sum_{n=1}^{N}
\sum_{t=1}^{T}
m_{n,t,i}
},
\]
\[
\text{covariates\_loss}_{val}
=
\sum_{i=1}^{C} loss_{L_i}.
\]

where $N$ denotes the total number of individuals in the validation set.

When training the outcome model, the loss function is defined as:

\[
\text{y\_loss}_{train}
=
\frac{
\displaystyle \sum_{b=1}^{B} 
\sum_{t=1}^{T}
m_{b,t}\,
\ell\!\left(\hat{y}_{b,t},\, y_{b,t}\right)
}{
\displaystyle \sum_{b=1}^{B} 
\sum_{t=1}^{T}
m_{b,t}
}.
\]

where $B$ denotes the number of individuals in each training batch, $T$ denotes the maximum sequence length. $m_{b, t}$ is a mask indicating if the outcome has missing value at time point $t$ for the $b$-th individual in the training batch. $\ell(\cdot)$ denotes the binary cross-entropy (BCE) loss for the outcome variable. 

On the validation set, the loss function for the outcome is defined as:

\[
\text{y\_loss}_{val}
=
\frac{
\displaystyle
\sum_{n=1}^{N}
\sum_{t=1}^{T}
m_{n,t}\,
\ell\!\left(\hat{y}_{n,t},\, y_{n,t}\right)
}{
\displaystyle
\sum_{n=1}^{N}
\sum_{t=1}^{T}
m_{n,t}
},
\]

where $N$ denotes the total number of individuals in the validation set.

\section{Hyperparameter search}
\label{app:space}
Table \ref{tab:hyperparams} summarizes the hyperparameter set for the LSTM used in modeling the conditional distributions of covariates and outcome, including each architectural and training-related hyperparameter and its corresponding search space. The architectural hyperparameters define the structure of the model, which includes the number of recurrent layers ($layers\_num$), the hidden state dimension of each LSTM layer ($hidden\_size$), and the feature dimension of each linear layer in the MLP decoder ($feature\_dim$). The training hyperparameters determine the learning process and include the learning rate ($lr$), batch size (
$batch\_size$), and weight decay ($weight\_decay$).

\begin{table}[htbp!]
\centering
\caption{Hyperparameter search space for the LSTMs in the covariate and outcome models.}
\label{tab:hyperparams}
\begin{tabular}{lll}
\toprule
\textbf{Hyperparameter} & \textbf{Covariate LSTM} & \textbf{Outcome LSTM} \\
\hline
$layers\_num$       & \{1, 2, 3\}              & \{1, 2, 3\}              \\
$hidden\_size$             & \{64, 128, 256\}         & \{64, 128, 256\}         \\
$dropout$            & uniform[0, 0.5]          & uniform[0, 0.5] \\
$feature\_dim$      & \{512, 1024\}             & \{512, 1024\}             \\
$lr$           & log-uniform[$10^{-5}$, $10^{-3}$] & log-uniform[$10^{-5}$, $10^{-3}$] \\
$batch\_size$              & \{512, 1024\}            & \{512, 1024\}            \\     
$weight\_decay$            & log-uniform[$10^{-6}$, $10^{-3}$] & log-uniform[$10^{-6}$, $10^{-3}$] \\
\bottomrule
\end{tabular}
\end{table}

\section{Data simulation}
\label{appendix:datagen}

Here we describe the data-generating process used in our simulation study, designed to resemble the structure and characteristics of real-world longitudinal HIV data.

\subsection{Simpler longitudinal data}

In the data generation, the time-varying covariates at time $k$ were generated to depend on an unmeasured covariate, baseline covariates, previous time-varying covariates at time $k-1$, previous treatment, and interactions between covariates. The outcome at time $k$ was generated from the current-time covariates and treatment.
The data-generating functions for each covariate and the outcome are shown below:

\subsubsection*{Unmeasured Covariate:}
\begin{itemize}
    \item $U \sim \mathcal{U}(0.0,\ 1.0)$
\end{itemize}

\subsubsection*{Baseline Covariates:}
\begin{itemize}
    \item \textit{Age} $\sim$ TruncatedNormal($\mu = 50$, $\sigma = 12$, $a = 18$, $b = 80$)
    \item \textit{Sex} $\sim$ Ber($p = 0.8$)
    \item \textit{Smoking} $\sim$ Multinomial$(1,\ [0.45,\ 0.4,\ 0.15])$
\end{itemize}

\subsubsection*{Treatment History Definition}
For covariate generation,
\begin{equation}
INSTI^{cov,lag}_{k} =
\begin{cases}
0, & k = 0, \\
INSTI_{k-1}, & k > 0.
\end{cases}
\end{equation}

\subsubsection*{CD4 Count:}
Initial value at time $k=-1$:
\begin{equation}
CD4_{k=-1} \sim \text{TruncatedNormal}(\mu=350, \sigma=30, a=300, b=400).
\end{equation}

For $0 \leq k \leq 5$:
\begin{equation}
\begin{aligned}
CD4_k &\sim \text{TruncatedNormal}(\mu, \sigma=40, a=300, b=1500) \\
\mu
&= 2 \cdot U + 5 \cdot Sex - 0.05 \cdot Age
   + 0 \cdot Age^2 + 0 \cdot Smoking \\
&\quad + 1.05 \cdot CD4_{k-1}
   + 0.0001 \cdot CD4_{k-1}^2
   - 0.2 \cdot RNA_{k-1} \\
&\quad - 0.1 \cdot HighBMI_{k-1}
   + 15 \cdot INSTI^{cov,lag}_{k} \\
&\quad + 0.1 \cdot HighBMI_{k-1} \cdot RNA_{k-1}
   + 0.08 \cdot HighBMI_{k-1} \cdot Sex \\
&\quad - 0.0001 \cdot CD4_{k-1} \cdot Age.
\end{aligned}
\end{equation}

For $k \geq 6$:
\begin{equation}
\begin{aligned}
CD4_k &\sim \text{TruncatedNormal}(\mu, \sigma=40, a=300, b=1500) \\
\mu
&= 2 \cdot U + 5 \cdot Sex - 0.05 \cdot Age
   + 0 \cdot Age^2 + 0 \cdot Smoking \\
&\quad + 1.03 \cdot CD4_{k-1}
   + 0.0001 \cdot CD4_{k-1}^2
   - 0.2 \cdot RNA_{k-1} \\
&\quad - 0.1 \cdot HighBMI_{k-1}
   + 15 \cdot INSTI^{cov,lag}_{k} \\
&\quad + 0.1 \cdot HighBMI_{k-1} \cdot RNA_{k-1}
   + 0.08 \cdot HighBMI_{k-1} \cdot Sex \\
&\quad - 0.0001 \cdot CD4_{k-1} \cdot Age.
\end{aligned}
\end{equation}

\subsubsection*{HIV RNA:}
Initial value at time $k=-1$:
\begin{equation}
RNA_{k=-1} \sim \text{TruncatedNormal}(\mu=210, \sigma=5, a=180, b=300).
\end{equation}

For $0 \leq k \leq 5$:
\begin{equation}
\begin{aligned}
RNA_k &\sim \text{TruncatedNormal}(\mu, \sigma=5, a=10, b=1000) \\
\mu
&= 1 \cdot U + 0.2 \cdot Sex + 0.01 \cdot Age
   + 0 \cdot Age^2 + 0 \cdot Smoking \\
&\quad - 0.0001 \cdot CD4_{k-1}
   - 0.000001 \cdot CD4_{k-1}^2
   + 0.94 \cdot RNA_{k-1} \\
&\quad + 0.3 \cdot HighBMI_{k-1}
   - 7 \cdot INSTI^{cov,lag}_{k} \\
&\quad + 0.0001 \cdot HighBMI_{k-1} \cdot RNA_{k-1}
   + 0.1 \cdot HighBMI_{k-1} \cdot Sex \\
&\quad + 0 \cdot CD4_{k-1} \cdot Age.
\end{aligned}
\end{equation}

For $k \geq 6$:
\begin{equation}
\begin{aligned}
RNA_k &\sim \text{TruncatedNormal}(\mu, \sigma=5, a=10, b=1000) \\
\mu
&= 1 \cdot U + 0.2 \cdot Sex + 0.01 \cdot Age
   + 0 \cdot Age^2 + 0 \cdot Smoking \\
&\quad - 0.0001 \cdot CD4_{k-1}
   - 0.000001 \cdot CD4_{k-1}^2
   + 0.96 \cdot RNA_{k-1} \\
&\quad + 0.3 \cdot HighBMI_{k-1}
   - 7 \cdot INSTI^{cov,lag}_{k} \\
&\quad + 0.0001 \cdot HighBMI_{k-1} \cdot RNA_{k-1}
   + 0.1 \cdot HighBMI_{k-1} \cdot Sex \\
&\quad + 0 \cdot CD4_{k-1} \cdot Age.
\end{aligned}
\end{equation}

\subsubsection*{BMI $>25$:}
Initial value at time $k=-1$:
\begin{equation}
HighBMI_{k=-1} \sim \text{Ber}(p=0.4).
\end{equation}

For $k \geq 0$:
\begin{equation}
\begin{aligned}
HighBMI_k &\sim \text{Ber}\left(\text{expit}(\mu)\right) \\
\mu
&= -8 - 2 \cdot U + 0.03 \cdot Sex
   + 0.01 \cdot Age + 0.0001 \cdot Age^2 \\
&\quad + 0.04 \cdot Smoking
   - 0.0001 \cdot CD4_{k-1}
   + 0 \cdot CD4_{k-1}^2 \\
&\quad + 0.001 \cdot RNA_{k-1}
   + 10 \cdot HighBMI_{k-1}
   + 5 \cdot INSTI^{cov,lag}_{k} \\
&\quad + 0.001 \cdot HighBMI_{k-1} \cdot RNA_{k-1}
   + 0.004 \cdot HighBMI_{k-1} \cdot Sex \\
&\quad + 0.00001 \cdot CD4_{k-1} \cdot Age.
\end{aligned}
\end{equation}

\subsubsection*{Treatment:}
Initial value at time $k=-1$:
\begin{equation}
INSTI_{k=-1} = 0.
\end{equation}

Under the natural course, for $k \geq 0$:
\begin{equation}
\begin{aligned}
INSTI_k &\sim \text{Ber}\left(\text{expit}(\mu)\right) \\
\mu
&= -4.5 + 0.5 \cdot Sex + 0.01 \cdot Age
   + 0.0001 \cdot Age^2 + 0.1 \cdot Smoking \\
&\quad + 0.001 \cdot CD4_k
   + 0 \cdot CD4_k^2
   + 0.01 \cdot RNA_k \\
&\quad - 7 \cdot HighBMI_k
   + 10 \cdot INSTI^{trt,lag}_{k} \\
&\quad + 0.0001 \cdot HighBMI_k \cdot RNA_k
   + 0.001 \cdot HighBMI_k \cdot Sex \\
&\quad + 0.00001 \cdot CD4_k \cdot Age.
\end{aligned}
\end{equation}

where
\begin{equation}
INSTI^{trt,lag}_{k} =
\begin{cases}
INSTI^{init}, & k = 0, \\
INSTI_{k-1}, & k > 0,
\end{cases}
\qquad
INSTI^{init} \sim \text{Ber}(0.5).
\end{equation}

\subsubsection*{Outcome:}

Initial value at time $k=-1$:
\begin{equation}
Y_{k=-1} = 0.
\end{equation}

For $k \geq 0$:
\begin{equation}
\begin{aligned}
Y_k &\sim \text{Ber}\left(\text{expit}(\mu)\right) \\
\mu
&= -0.002 \cdot U
   + 0.0001 \cdot Sex
   + 0.00015 \cdot Age
   + 0.0025 \cdot Smoking \\
&\quad - 0.025 \cdot CD4_k
   + 0.03 \cdot RNA_k
   + 0.000004 \cdot RNA_k^2 \\
&\quad + 0.01 \cdot HighBMI_k
   - 0.12 \cdot INSTI_k.
\end{aligned}
\end{equation}

\subsubsection*{Interventions:}
Under interventions, treatment was assigned as follows:
\begin{itemize}
\def\labelitemi{$\bullet$}
    \item Natural course: $INSTI_k$ was generated from the treatment model above.
    \item Always treat: $INSTI_k = 1$ for all $k \geq 0$.
    \item Never treat: $INSTI_k = 0$ for all $k \geq 0$.
    \item Dynamic intervention: $INSTI_k=1$ if treatment had already been initiated or if $RNA_k>200$; otherwise $INSTI_k=0$.
\end{itemize}

\subsection{More complex longitudinal data}

In the data generation, the time-varying covariates and the outcome were generated conditional on the unmeasured covariate, baseline covariates, the entire history of the time-varying covariates, and interactions between covariates. We further generated 30 pre-baseline time points ($k=-30, \dots, -1$) for all time-varying covariates to simulate pre-baseline history. We generated the values of the continuous covariate CD4 at the pre-baseline time points by decreasing the values by 0.5\% per month and the continuous covariate HIV-RNA by increasing the values by 1\% per month. The pre-baseline values for the binary covariate BMI were set to the baseline value (value at time $k=0$), and the pre-baseline values for the binary treatment INSTI use were set to 0 to reflect the absence of treatment before baseline. The data-generating functions for each covariate and
the outcome are shown below:

\subsubsection*{Unmeasured Covariate:}
\begin{itemize}
    \item $U \sim \mathcal{U}(0.0,\ 1.0)$
\end{itemize}

\subsubsection*{Baseline Covariates:}
\begin{itemize}
    \item \textit{Age} $\sim$ TruncatedNormal($\mu = 50$, $\sigma = 12$, $a = 18$, $b = 80$)
    \item \textit{Sex} $\sim$ Ber($p = 0.8$)
    \item \textit{Smoking} $\sim$ Multinomial$(1,\ [0.45,\ 0.4,\ 0.15])$
\end{itemize}

\subsubsection*{Treatment History Definitions}
For covariate generation,
\begin{equation}
INSTI^{cov,lag}_{k} =
\begin{cases}
0, & k = 0, \\
INSTI_{k-1}, & k > 0.
\end{cases}
\end{equation}

\begin{equation}
\overline{INSTI}^{hist}_{k} =
\left\{
\begin{alignedat}{2}
&0,
&&\qquad k = 0, \\[6pt]
&\displaystyle \frac{1}{k+30}\sum_{j=-30}^{k-1} INSTI_j,
&&\qquad 1 \leq k \leq 30, \\[10pt]
&\displaystyle \frac{1}{60}\sum_{j=k-60}^{k-1} INSTI_j,
&&\qquad k > 30.
\end{alignedat}
\right.
\end{equation}

\subsubsection*{CD4 Count:}
Pre-baseline values:

\begin{equation}
    CD4_k = CD4^{init} \cdot \frac{1}{0.995^{|k|}},
\qquad k=-30,\ldots,-1,
\end{equation}

where
\begin{equation}
    CD4^{init} \sim \text{TruncatedNormal}(\mu=350, \sigma=30, a=300, b=400).
\end{equation}

For $0 \leq k < 6$:
\begin{equation}
\begin{aligned}
CD4_k &\sim \text{TruncatedNormal}(\mu, \sigma=40, a=300, b=1500) \\
\mu
&= -2 \cdot U + 0.1 \cdot Sex - 1 \cdot Age
   - 0.005 \cdot Age^2 + 0.3 \cdot Smoking \\
&\quad + 1.03 \cdot CD4_{k-1}
   - 0.00005 \cdot CD4_{k-1}^2
   + 0.15 \cdot \overline{CD4}_{k-6:k-1} \\
&\quad + 0.15 \cdot \overline{CD4}_{k-24:k-8}
   + 0.1 \cdot \overline{CD4}_{-30:k-26}
   - 0.08 \cdot RNA_{k-1} \\
&\quad - 0.05 \cdot \overline{RNA}_{k-6:k-1}
   - 0.03 \cdot \overline{RNA}_{k-24:k-8}
   - 0.01 \cdot \overline{RNA}_{-30:k-26} \\
&\quad - 0.1 \cdot HighBMI_{k-1}
   - 0.08 \cdot \overline{HighBMI}_{k-6:k-1} \\
&\quad - 0.04 \cdot \overline{HighBMI}_{k-24:k-8}
   - 0.02 \cdot \overline{HighBMI}_{-30:k-26} \\
&\quad + 25 \cdot INSTI^{cov,lag}_{k}
   + 6.2 \cdot \overline{INSTI}^{hist}_{k} \\
&\quad + 0.05 \cdot HighBMI_{k-1} \cdot RNA_{k-1}
   + 0.04 \cdot HighBMI_{k-1} \cdot Sex \\
&\quad - 0.0005 \cdot CD4_{k-1} \cdot Age.
\end{aligned}
\end{equation}

For $k \geq 6$:
\begin{equation}
\begin{aligned}
CD4_k &\sim \text{TruncatedNormal}(\mu, \sigma=40, a=300, b=1500) \\
\mu
&= -2 \cdot U + 0.1 \cdot Sex - 1 \cdot Age
   - 0.005 \cdot Age^2 + 0.3 \cdot Smoking \\
&\quad + 1.02 \cdot CD4_{k-1}
   - 0.00005 \cdot CD4_{k-1}^2
   + 0.2 \cdot \overline{CD4}_{k-6:k-1} \\
&\quad + 0.1 \cdot \overline{CD4}_{k-24:k-8}
   + 0.05 \cdot \overline{CD4}_{-30:k-26}
   - 0.08 \cdot RNA_{k-1} \\
&\quad - 0.05 \cdot \overline{RNA}_{k-6:k-1}
   - 0.03 \cdot \overline{RNA}_{k-24:k-8}
   - 0.01 \cdot \overline{RNA}_{-30:k-26} \\
&\quad - 0.1 \cdot HighBMI_{k-1}
   - 0.08 \cdot \overline{HighBMI}_{k-6:k-1} \\
&\quad - 0.04 \cdot \overline{HighBMI}_{k-24:k-8}
   - 0.02 \cdot \overline{HighBMI}_{-30:k-26} \\
&\quad + 25 \cdot INSTI^{cov,lag}_{k}
   + 6.2 \cdot \overline{INSTI}^{hist}_{k} \\
&\quad + 0.05 \cdot HighBMI_{k-1} \cdot RNA_{k-1}
   + 0.04 \cdot HighBMI_{k-1} \cdot Sex \\
&\quad - 0.0005 \cdot CD4_{k-1} \cdot Age.
\end{aligned}
\end{equation}

\subsubsection*{HIV RNA:}
Pre-baseline values:

\begin{equation}
    RNA_k = RNA^{init} \cdot \frac{1}{1.01^{|k|}},
\qquad k=-30,\ldots,-1,
\end{equation}

where
\begin{equation}
    RNA^{init} \sim \text{TruncatedNormal}(\mu=200, \sigma=5, a=180, b=230).
\end{equation}

For $0 \leq k < 6$:
\begin{equation}
\begin{aligned}
RNA_k &\sim \text{TruncatedNormal}(\mu, \sigma=10, a=10, b=1000) \\
\mu
&= 2 \cdot U + 1 \cdot Sex + 1.5 \cdot Age
   + 0.006 \cdot Age^2 + 0.5 \cdot Smoking \\
&\quad - 0.1 \cdot CD4_{k-1}
   - 0.00001 \cdot CD4_{k-1}^2
   - 0.2 \cdot \overline{CD4}_{k-6:k-1} \\
&\quad - 0.005 \cdot \overline{CD4}_{k-24:k-8}
   - 0.001 \cdot \overline{CD4}_{-30:k-26}
   + 0.93 \cdot RNA_{k-1} \\
&\quad + 0.05 \cdot \overline{RNA}_{k-6:k-1}
   + 0.05 \cdot \overline{RNA}_{k-24:k-8}
   + 0.02 \cdot \overline{RNA}_{-30:k-26} \\
&\quad + 0.3 \cdot HighBMI_{k-1}
   + 0.1 \cdot \overline{HighBMI}_{k-6:k-1} \\
&\quad + 0.06 \cdot \overline{HighBMI}_{k-24:k-8}
   + 0.03 \cdot \overline{HighBMI}_{-30:k-26} \\
&\quad - 25 \cdot INSTI^{cov,lag}_{k}
   - 1.52 \cdot \overline{INSTI}^{hist}_{k} \\
&\quad + 0.04 \cdot HighBMI_{k-1} \cdot RNA_{k-1}
   + 0.05 \cdot HighBMI_{k-1} \cdot Sex \\
&\quad - 0.00005 \cdot CD4_{k-1} \cdot Age.
\end{aligned}
\end{equation}

For $k \geq 6$:
\begin{equation}
\begin{aligned}
RNA_k &\sim \text{TruncatedNormal}(\mu, \sigma=10, a=10, b=1000) \\
\mu
&= 2 \cdot U + 1 \cdot Sex + 1.5 \cdot Age
   + 0.006 \cdot Age^2 + 0.5 \cdot Smoking \\
&\quad - 0.1 \cdot CD4_{k-1}
   - 0.00001 \cdot CD4_{k-1}^2
   - 0.2 \cdot \overline{CD4}_{k-6:k-1} \\
&\quad - 0.005 \cdot \overline{CD4}_{k-24:k-8}
   - 0.001 \cdot \overline{CD4}_{-30:k-26}
   + 0.93 \cdot RNA_{k-1} \\
&\quad + 0.05 \cdot \overline{RNA}_{k-6:k-1}
   + 0.03 \cdot \overline{RNA}_{k-24:k-8}
   + 0.03 \cdot \overline{RNA}_{-30:k-26} \\
&\quad + 0.3 \cdot HighBMI_{k-1}
   + 0.1 \cdot \overline{HighBMI}_{k-6:k-1} \\
&\quad + 0.06 \cdot \overline{HighBMI}_{k-24:k-8}
   + 0.03 \cdot \overline{HighBMI}_{-30:k-26} \\
&\quad - 25 \cdot INSTI^{cov,lag}_{k}
   - 1.52 \cdot \overline{INSTI}^{hist}_{k} \\
&\quad + 0.04 \cdot HighBMI_{k-1} \cdot RNA_{k-1}
   + 0.05 \cdot HighBMI_{k-1} \cdot Sex \\
&\quad - 0.00005 \cdot CD4_{k-1} \cdot Age.
\end{aligned}
\end{equation}

\subsubsection*{BMI $>25$:}
Pre-baseline values:
\begin{equation}
HighBMI_k = HighBMI^{init},
\qquad k=-30,\ldots,-1,
\qquad HighBMI^{init}\sim \text{Ber}(0.4).
\end{equation}

For $k \geq 0$:
\begin{equation}
\begin{aligned}
HighBMI_k &\sim \text{Ber}\left(\text{expit}(\mu)\right) \\
\mu
&= -6.5 - 1 \cdot U - 0.6 \cdot Sex
   + 0.01 \cdot Age + 0.0001 \cdot Age^2 \\
&\quad - 0.04 \cdot Smoking
   - 0.0001 \cdot CD4_{k-1}
   + 0.000001 \cdot CD4_{k-1}^2 \\
&\quad - 0.001 \cdot \overline{CD4}_{k-6:k-1}
   - 0.0001 \cdot \overline{CD4}_{k-24:k-8}
   - 0.001 \cdot \overline{CD4}_{-30:k-26} \\
&\quad + 0.01 \cdot RNA_{k-1}
   + 0.01 \cdot \overline{RNA}_{k-6:k-1}
   + 0.007 \cdot \overline{RNA}_{k-24:k-8} \\
&\quad + 0.006 \cdot \overline{RNA}_{-30:k-26}
   + 4.5 \cdot HighBMI_{k-1}
   + 3 \cdot \overline{HighBMI}_{k-6:k-1} \\
&\quad + 1.6 \cdot \overline{HighBMI}_{k-24:k-8}
   + 1 \cdot \overline{HighBMI}_{-30:k-26} \\
&\quad + 2 \cdot INSTI^{cov,lag}_{k}
   + 1 \cdot \overline{INSTI}^{hist}_{k} \\
&\quad + 0 \cdot HighBMI_{k-1} \cdot RNA_{k-1}
   + 0 \cdot HighBMI_{k-1} \cdot Sex \\
&\quad + 0 \cdot HighBMI_{k-1} \cdot Age.
\end{aligned}
\end{equation}

\subsubsection*{Treatment:}
Pre-baseline values:
\begin{equation}
INSTI_k = 0,
\qquad k=-30,\ldots,-1.
\end{equation}

Under the natural course, for $k \geq 0$:
\begin{equation}
\begin{aligned}
INSTI_k &\sim \text{Ber}\left(\text{expit}(\mu)\right) \\
\mu
&= -4 + 0.5 \cdot Sex + 0.05 \cdot Age
   + 0.00005 \cdot Age^2 + 0.2 \cdot Smoking \\
&\quad - 0.004 \cdot CD4_k
   - 0.0000001 \cdot CD4_k^2
   - 0.0001 \cdot \overline{CD4}_{k-6:k-1} \\
&\quad + 0 \cdot \overline{CD4}_{k-24:k-7}
   + 0 \cdot \overline{CD4}_{-30:k-25}
   + 0.001 \cdot RNA_k \\
&\quad + 0.0003 \cdot \overline{RNA}_{k-6:k-1}
   + 0 \cdot \overline{RNA}_{k-24:k-7}
   + 0 \cdot \overline{RNA}_{-30:k-25} \\
&\quad - 3 \cdot HighBMI_k
   - 2 \cdot \overline{HighBMI}_{k-6:k-1}
   - 1.3 \cdot \overline{HighBMI}_{k-24:k-7} \\
&\quad - 0.8 \cdot \overline{HighBMI}_{-30:k-25}
   + 10.5 \cdot INSTI^{trt,lag}_{k}
   + 6.8 \cdot \overline{INSTI}^{hist}_{k} \\
&\quad + 0 \cdot HighBMI_k \cdot RNA_k
   + 0 \cdot HighBMI_k \cdot Sex
   + 0 \cdot HighBMI_k \cdot Age.
\end{aligned}
\end{equation}

where
\begin{equation}
INSTI^{trt,lag}_{k} =
\begin{cases}
INSTI^{init}, & k = 0, \\
INSTI_{k-1}, & k > 0,
\end{cases}
\qquad
INSTI^{init} \sim \text{Ber}(0.5).
\end{equation}

\subsubsection*{Outcome:}
Pre-baseline outcome values:
\begin{equation}
Y_k = 0,
\qquad k=-30,\ldots,-1.
\end{equation}

For $k \geq 0$:
\begin{equation}
\begin{aligned}
Y_k &\sim \text{Ber}\left(\text{expit}(\mu)\right) \\
\mu
&= -0.05 \cdot U + 0.007 \cdot Sex + 0.02 \cdot Age
   + 0.00000005 \cdot Age^2 \\
&\quad + 0.03 \cdot Smoking
   - 0.009 \cdot CD4_k
   - 0.008 \cdot \overline{CD4}_{k-6:k-1} \\
&\quad - 0.006 \cdot \overline{CD4}_{k-24:k-8}
   - 0.004 \cdot \overline{CD4}_{-30:k-26}
   + 0.015 \cdot RNA_k \\
&\quad + 0.0000004 \cdot RNA_k^2
   + 0.01 \cdot \overline{RNA}_{k-6:k-1}
   + 0.005 \cdot \overline{RNA}_{k-24:k-8} \\
&\quad + 0.002 \cdot \overline{RNA}_{-30:k-26}
   + 0.14 \cdot HighBMI_k
   + 0.11 \cdot \overline{HighBMI}_{k-6:k-1} \\
&\quad + 0.08 \cdot \overline{HighBMI}_{k-24:k-8}
   + 0.06 \cdot \overline{HighBMI}_{-30:k-26} \\
&\quad + 0.009 \cdot INSTI_k
   - 0.008 \cdot \overline{INSTI}^{hist}_{k}.
\end{aligned}
\end{equation}

\subsubsection*{Interventions:}
Under interventions, treatment was assigned as follows:
\begin{itemize}
\def\labelitemi{$\bullet$}
    \item Natural course: $INSTI_k$ was generated from the treatment model.
    \item Always treat: $INSTI_k=1$ for all $k\ge 0$.
    \item Never treat: $INSTI_k=0$ for all $k\ge 0$.
    \item Dynamic intervention: $INSTI_k=1$ if treatment had already been initiated or if $RNA_k>200$; otherwise $INSTI_k=0$.
\end{itemize}

\section{Model parameterization in the parametric g-formula estimator}
\label{supp:model_spec}

Here, we describe the parameterization of each model used in the parametric g-formula estimator. For both the simple and complex data scenarios, we consider two parameterizations: one that more closely approximates the true data-generating process and another that deviates further from it.

\subsection{Simpler longitudinal data}

\textbf{Larger model misspecification:}

At baseline time $k=0$, the empirical distribution of each time-varying covariate was used. For times $k=1,...,9$, we specified a pooled parametric model for the conditional distribution of each time-varying covariate given the observed history, with the same model and parameters applied across time points. Similarly, a pooled parametric model was specified for the conditional distribution of the outcome given the observed history at times $k=0,...,9$. Time-invariant baseline covariates were included as predictors in all conditional models for time-varying covariates and the outcome. No parametric model was specified for the marginal distribution of baseline covariates. This specification applies to all parameterization settings considered below.

\[
\begin{aligned}
\text{RNA}_k 
&\sim \mathcal{N}(\mu_k, \sigma_{\text{RNA}}^2),
\\[6pt]
\mu_k
&= \beta_0
+ \beta_1 \, \text{CD4}_{k-1}
+ \beta_2 \, \text{RNA}_{k-1}
+ \beta_3 \, \text{HighBMI}_{k-1}
+ \beta_4 \, \text{INSTI}_{k-1}
+ \beta_5 \, \text{Sex}
\\
&\quad
+ \beta_6 \, \text{Age}
+ \beta_7 \, \text{Smoking}
+ \beta_8 \, k .
\end{aligned}
\]

\[
\begin{aligned}
\text{CD4}_k 
&\sim \mathcal{N}(\mu_k, \sigma_{\text{CD4}}^2),
\\[6pt]
\mu_k
&= \gamma_0
+ \gamma_1 \, \text{CD4}_{k-1}
+ \gamma_2 \, \text{RNA}_{k-1}
+ \gamma_3 \, \text{HighBMI}_{k-1}
+ \gamma_4 \, \text{INSTI}_{k-1}
+ \gamma_5 \, \text{Sex}
\\
&\quad
+ \gamma_6 \, \text{Age}
+ \gamma_7 \, \text{Smoking}
+ \gamma_8 \, k .
\end{aligned}
\]

\[
\begin{aligned}
\text{HighBMI}_k
&\sim \text{Bernoulli}\!\left(
\text{logit}^{-1}(\mu_k)
\right),
\\[6pt]
\mu_k
&= \delta_0
+ \delta_1 \, \text{CD4}_{k-1}
+ \delta_2 \, \text{RNA}_{k-1}
+ \delta_3 \, \text{HighBMI}_{k-1}
+ \delta_4 \, \text{INSTI}_{k-1}
+ \delta_5 \, \text{Sex}
\\
&\quad
+ \delta_6 \, \text{Age}
+ \delta_7 \, \text{Smoking}
+ \delta_8 \, k .
\end{aligned}
\]

\[
\begin{aligned}
\text{INSTI}_k
&\sim \text{Bernoulli}\!\left(
\text{logit}^{-1}(\mu_k)
\right),
\\[6pt]
\mu_k
&= \eta_0
+ \eta_1 \, \text{CD4}_{k-1}
+ \eta_2 \, \text{RNA}_{k-1}
+ \eta_3 \, \text{HighBMI}_{k-1}
+ \eta_4 \, \text{INSTI}_{k-1}
\\
&\quad
+ \eta_5 \, \text{Sex}
+ \eta_6 \, \text{Age}
+ \eta_7 \, \text{Smoking}
+ \eta_8 \, k .
\end{aligned}
\]

\[
\begin{aligned}
\text{Event}_k
&\sim \text{Bernoulli}\!\left(
\text{logit}^{-1}(\mu_k)
\right),
\\[6pt]
\mu_k
&= \alpha_0
+ \alpha_1 \, \text{CD4}_k
+ \alpha_2 \, \text{RNA}_k
+ \alpha_3 \, \text{HighBMI}_k
+ \alpha_4 \, \text{INSTI}_k
+ \alpha_5 \, \text{Sex}
\\
&\quad
+ \alpha_6 \, \text{Age}
+ \alpha_7 \, \text{Smoking}
+ \alpha_8 \, k .
\end{aligned}
\]

\vspace{5mm}

\textbf{Smaller model misspecification:}

\[
\begin{aligned}
\text{CD4}_k
&\sim \mathcal{N}(\mu_k, \sigma_{\text{CD4}}^2),
\\[6pt]
\mu_k
&= \beta_0
+ \beta_1 \, \text{Sex}
+ \beta_2 \, \text{Age}
+ \beta_3 \, \text{CD4}_{k-1}
+ \beta_4 \, \text{CD4}_{k-1}^2
+ \beta_5 \, \text{RNA}_{k-1}
+ \beta_6 \, \text{HighBMI}_{k-1}
+ \beta_7 \, \text{INSTI}_{k-1}
\\
&\quad
+ \beta_8 \, \text{HighBMI}_{k-1} \cdot \text{RNA}_{k-1}
+ \beta_9 \, \text{HighBMI}_{k-1} \cdot \text{Sex}
+ \beta_{10} \, \text{CD4}_{k-1} \cdot \text{Age}
+ \beta_{11} \, \text{CD4}_{k-1} \cdot I(k > 5).
\end{aligned}
\]

\[
\begin{aligned}
\text{RNA}_k
&\sim \mathcal{N}(\mu_k, \sigma_{\text{RNA}}^2),
\\[6pt]
\mu_k
&= \gamma_0
+ \gamma_1 \, \text{Sex}
+ \gamma_2 \, \text{Age}
+ \gamma_3 \, \text{CD4}_{k-1}
+ \gamma_4 \, \text{CD4}_{k-1}^2
+ \gamma_5 \, \text{RNA}_{k-1}
+ \gamma_6 \, \text{HighBMI}_{k-1}
+ \gamma_7 \, \text{INSTI}_{k-1}
\\
&\quad
+ \gamma_8 \, \text{HighBMI}_{k-1} \cdot \text{RNA}_{k-1}
+ \gamma_9 \, \text{HighBMI}_{k-1} \cdot \text{Sex}
+ \gamma_{10} \, \text{RNA}_{k-1} \cdot I(k > 5).
\end{aligned}
\]

\[
\begin{aligned}
\text{HighBMI}_k
&\sim \text{Bernoulli}\!\left(
\text{logit}^{-1}(\mu_k)
\right),
\\[6pt]
\mu_k
&= \delta_0
+ \delta_1 \, \text{Sex}
+ \delta_2 \, \text{Age}
+ \delta_3 \, \text{Age}^2
+ \delta_4 \, \text{Smoking}
+ \delta_5 \, \text{CD4}_{k-1}
+ \delta_6 \, \text{RNA}_{k-1}
+ \delta_7 \, \text{HighBMI}_{k-1}
\\
&\quad
+ \delta_8 \, \text{INSTI}_{k-1}
+ \delta_9 \, \text{HighBMI}_{k-1} \cdot \text{RNA}_{k-1}
+ \delta_{10} \, \text{HighBMI}_{k-1} \cdot \text{Sex}
+ \delta_{11} \, \text{CD4}_{k-1} \cdot \text{Age}
+ \delta_{12} \, k .
\end{aligned}
\]

\[
\begin{aligned}
\text{INSTI}_k
&\sim \text{Bernoulli}\!\left(
\text{logit}^{-1}(\mu_k)
\right),
\\[6pt]
\mu_k
&= \eta_0
+ \eta_1 \, \text{Sex}
+ \eta_2 \, \text{Age}
+ \eta_3 \, \text{Age}^2
+ \eta_4 \, \text{Smoking}
+ \eta_5 \, \text{CD4}_k
+ \eta_6 \, \text{RNA}_k
+ \eta_7 \, \text{HighBMI}_k
\\
&\quad
+ \eta_8 \, \text{INSTI}_{k-1}
+ \eta_9 \, \text{HighBMI}_k \cdot \text{RNA}_k
+ \eta_{10} \, \text{HighBMI}_k \cdot \text{Sex}
+ \eta_{11} \, \text{CD4}_k \cdot \text{Age}
+ \eta_{12} \, k .
\end{aligned}
\]

\[
\begin{aligned}
\text{Event}_k
&\sim \text{Bernoulli}\!\left(
\text{logit}^{-1}(\mu_k)
\right),
\\[6pt]
\mu_k
&= \alpha_0
+ \alpha_1 \, \text{CD4}_k
+ \alpha_2 \, \text{RNA}_k
+ \alpha_3 \, \text{RNA}_k^2
+ \alpha_4 \, \text{HighBMI}_k
+ \alpha_5 \, \text{INSTI}_k
+ \alpha_6 \, \text{Sex}
\\
&\quad
+ \alpha_7 \, \text{Age}
+ \alpha_8 \, \text{Smoking}
+ \alpha_9 \, k .
\end{aligned}
\]

\subsection{More complex longitudinal data}

\textbf{Larger model misspecification:} same as in the simpler HIV data scenario.

\textbf{Smaller model misspecification:}

\[
\begin{aligned}
\text{CD4}_k 
&\sim \mathcal{N}(\mu_k, \sigma_{\text{CD4}}^2),
\\[6pt]
\mu_k
&= \beta_0
+ \beta_1 \text{Sex}
+ \beta_2 \text{Age}
+ \beta_3 \text{Age}^2
+ \beta_4 \text{Smoking}
+ \beta_5 \text{CD4}_{k-1}
+ \beta_6 \text{CD4}_{k-1}^2
+ \beta_7 \overline{\text{CD4}}_{0:k-1}
\\
&\quad
+ \beta_{8} \text{RNA}_{k-1}
+ \beta_{9} \overline{\text{RNA}}_{0:k-1}
+ \beta_{10} \text{HighBMI}_{k-1}
+ \beta_{11} \overline{\text{HighBMI}}_{0:k-1}
+ \beta_{12} \text{INSTI}_{k-1}
\\
&\quad
+ \beta_{13} \overline{\text{INSTI}}_{0:k-1}
+ \beta_{14} \text{HighBMI}_{k-1} \cdot \text{RNA}_{k-1}
+ \beta_{15} \text{HighBMI}_{k-1} \cdot \text{Sex}
\\
&\quad
+ \beta_{16} \text{CD4}_{k-1} \cdot \text{Age}
+ \beta_{17} \text{CD4}_{k-1} \cdot I(k>5)
+ \beta_{18} \overline{\text{CD4}}_{0:k-1} \cdot  I(k>5)
\end{aligned}
\]

\[
\begin{aligned}
\text{RNA}_k 
&\sim \mathcal{N}(\mu_k, \sigma_{\text{RNA}}^2),
\\[6pt]
\mu_k
&= \gamma_0
+ \gamma_1 \text{Sex}
+ \gamma_2 \text{Age}
+ \gamma_3 \text{Age}^2
+ \gamma_4 \text{Smoking}
+ \gamma_5 \text{CD4}_{k-1}
+ \gamma_6 \text{CD4}_{k-1}^2
+ \gamma_7 \overline{\text{CD4}}_{0:k-1}
\\
&\quad
+ \gamma_{8} \text{RNA}_{k-1}
+ \gamma_{9} \overline{\text{RNA}}_{0:k-1}
+ \gamma_{10} \text{HighBMI}_{k-1}
+ \gamma_{11} \overline{\text{HighBMI}}_{0:k-1}
+ \gamma_{12} \text{INSTI}_{k-1}
\\
&\quad
+ \gamma_{13} \overline{\text{INSTI}}_{0:k-1}
+ \gamma_{14} \text{HighBMI}_{k-1} \cdot \text{RNA}_{k-1}
+ \gamma_{15} \text{HighBMI}_{k-1} \cdot \text{Sex}
\\
&\quad
+ \gamma_{16} \text{CD4}_{k-1} \cdot \text{Age}
+ \gamma_{17} \overline{\text{RNA}}_{0:k-1} \cdot I(k>5)
\end{aligned}
\]

\[
\begin{aligned}
\text{HighBMI}_k
&\sim \text{Bernoulli}\!\left(
\text{logit}^{-1}(\mu_k)
\right),
\\[6pt]
\mu_k
&= \delta_0
+ \delta_1 \text{Sex}
+ \delta_2 \text{Age}
+ \delta_3 \text{Age}^2
+ \delta_4 \text{Smoking}
+ \delta_5 \text{CD4}_{k-1}
+ \delta_6 \text{CD4}_{k-1}^2
+ \delta_7 \overline{\text{CD4}}_{0:k-1}
\\
&\quad
+ \delta_{8} \text{RNA}_{k-1}
+ \delta_{9} \overline{\text{RNA}}_{0:k-1}
+ \delta_{10} \text{HighBMI}_{k-1}
+ \delta_{11} \overline{\text{HighBMI}}_{0:k-1}
\\
&\quad
+ \delta_{12} \text{INSTI}_{k-1}
+ \delta_{13} \overline{\text{INSTI}}_{0:k-1}
+ \delta_{14} k .
\end{aligned}
\]

\[
\begin{aligned}
\text{INSTI}_k
&\sim \text{Bernoulli}\!\left(
\text{logit}^{-1}(\mu_k)
\right),
\\[6pt]
\mu_k
&= \eta_0
+ \eta_1 \text{Sex}
+ \eta_2 \text{Age}
+ \eta_3 \text{Age}^2
+ \eta_4 \text{Smoking}
+ \eta_5 \text{CD4}_{k}
+ \eta_6 \text{CD4}_{k}^2
+ \eta_7 \overline{\text{CD4}}_{0:k-1}
\\
&\quad
+ \eta_{10} \text{RNA}_{k}
+ \eta_{11} \overline{\text{RNA}}_{0:k-1}
+ \eta_{14} \text{HighBMI}_{k}
+ \eta_{15} \overline{\text{HighBMI}}_{0:k-1}
\\
&\quad
+ \eta_{18} \text{INSTI}_{k-1}
+ \eta_{19} \overline{\text{INSTI}}_{0:k-1}
+ \eta_{20} k .
\end{aligned}
\]

\[
\begin{aligned}
\text{Event}_k
&\sim \text{Bernoulli}\!\left(
\text{logit}^{-1}(\mu_k)
\right),
\\[6pt]
\mu_k
&= \alpha_0
+ \alpha_1 \text{Sex}
+ \alpha_2 \text{Age}
+ \alpha_3 \text{Age}^2
+ \alpha_4 \text{Smoking}
+ \alpha_5 \text{CD4}_k
+ \alpha_6 \overline{\text{CD4}}_{0:k-1}
\\
&\quad
+ \alpha_9 \text{RNA}_k
+ \alpha_{10} \text{RNA}_k^2
+ \alpha_{11} \overline{\text{RNA}}_{0:k-1}
+ \alpha_{14} \text{HighBMI}_k
+ \alpha_{15} \overline{\text{HighBMI}}_{0:k-1}
\\
&\quad
+ \alpha_{18} \text{INSTI}_k
+ \alpha_{19} \overline{\text{INSTI}}_{0:k-1}
+ \alpha_{20} k .
\end{aligned}
\]

Here, $\overline{Var}_{a:b}$ denotes the average of the values of variable $Var$ between time points $a$ and $b$, inclusive.

\newpage

\section{Quantitative results for risk differences and risk ratios}
\label{supp:risk_ratio_difference}

\begin{table}[!htbp]
\centering
\caption{\textbf{Simple longitudinal HIV data.} Risk differences and risk ratios relative to the natural course under three interventions.}
\label{tab:simple_rd_rr_all_interventions_coverage_truth}
\small
\setlength{\tabcolsep}{3pt}
\renewcommand{\arraystretch}{1.12}
\resizebox{\textwidth}{!}{%
\begin{tabular}{clcccc cccc}
\toprule
Sample size & Estimator
& \multicolumn{4}{c}{Risk difference ($\%$)}
& \multicolumn{4}{c}{Risk ratio} \\
\cmidrule(lr){3-6}\cmidrule(lr){7-10}
& & Estimate & Std.\ error & 95\% CI & Coverage ($\%$) & Estimate & Std.\ error & 95\% CI & Coverage ($\%$) \\
\hline
\multicolumn{10}{l}{\textbf{Always treat} (true risk difference = -2.77\%, true risk ratio = 0.81)} \\
\addlinespace[2pt]
\multirow{4}{*}{1,000}
& LSTM with sample splitting
& -2.74 & 0.029 & (-2.800, -2.685) & 93.8
& 0.81 & 0.002 & (0.806, 0.814) & 94.5 \\
& LSTM without sample splitting
& -2.86 & 0.031 & (-2.919, -2.797) & 92.1
& 0.80 & 0.002 & (0.795, 0.803) & 91.7 \\
& Parametric with larger misspecification
& -3.11 & 0.027 & (-3.159, -3.052) & 93.4
& 0.80 & 0.002 & (0.794, 0.801) & 94.3 \\
& Parametric with smaller misspecification
& -2.93 & 0.027 & (-2.984, -2.880) & 94.6
& 0.80 & 0.002 & (0.801, 0.808) & 95.3 \\

\addlinespace[3pt]
\multirow{4}{*}{10,000}
& LSTM with sample splitting
& -2.66 & 0.010 & (-2.679, -2.639) & 93.0
& 0.81 & 0.001 & (0.811, 0.814) & 93.2 \\
& LSTM without sample splitting
& -2.64 & 0.010 & (-2.662, -2.623) & 94.1
& 0.81 & 0.001 & (0.812, 0.815) & 94.5 \\
& Parametric with larger misspecification
& -3.13 & 0.009 & (-3.147, -3.112) & 76.0
& 0.80 & 0.001 & (0.795, 0.797) & 89.1 \\
& Parametric with smaller misspecification
& -2.95 & 0.009 & (-2.968, -2.935) & 91.6
& 0.80 & 0.001 & (0.802, 0.804) & 94.0 \\
\hline
\multicolumn{10}{l}{\textbf{Never treat} (true risk difference = 6.36\%, true risk ratio = 1.44)} \\
\addlinespace[2pt]
\multirow{4}{*}{1,000}
& LSTM with sample splitting
& 6.26 & 0.074 & (6.118, 6.409) & 93.1
& 1.43 & 0.005 & (1.424, 1.444) & 93.9 \\
& LSTM without sample splitting
& 6.52 & 0.077 & (6.372, 6.673) & 90.4
& 1.46 & 0.005 & (1.449, 1.469) & 91.7 \\
& Parametric with larger misspecification
& 7.20 & 0.067 & (7.065, 7.327) & 93.0
& 1.47 & 0.004 & (1.461, 1.478) & 94.0 \\
& Parametric with smaller misspecification
& 6.84 & 0.064 & (6.710, 6.962) & 94.2
& 1.46 & 0.004 & (1.448, 1.465) & 95.0 \\

\addlinespace[3pt]
\multirow{4}{*}{10,000}
& LSTM with sample splitting
& 6.06 & 0.025 & (6.008, 6.106) & 93.3
& 1.43 & 0.002 & (1.424, 1.430) & 94.4 \\
& LSTM without sample splitting
& 6.10 & 0.026 & (6.046, 6.149) & 94.3
& 1.43 & 0.002 & (1.426, 1.433) & 95.3 \\
& Parametric with larger misspecification
& 7.24 & 0.021 & (7.200, 7.283) & 75.1
& 1.47 & 0.001 & (1.470, 1.475) & 89.2 \\
& Parametric with smaller misspecification
& 6.87 & 0.020 & (6.835, 6.914) & 90.2
& 1.46 & 0.001 & (1.456, 1.461) & 94.1 \\
\hline
\multicolumn{10}{l}{\textbf{Dynamic intervention} (true risk difference = 2.01\%, true risk ratio = 1.14)} \\
\addlinespace[2pt]
\multirow{4}{*}{1,000}
& LSTM with sample splitting
& 2.26 & 0.030 & (2.200, 2.317) & 95.8
& 1.16 & 0.002 & (1.153, 1.161) & 96.4 \\
& LSTM without sample splitting
& 2.29 & 0.032 & (2.222, 2.348) & 93.4
& 1.16 & 0.002 & (1.157, 1.166) & 93.9 \\
& Parametric with larger misspecification
& 2.34 & 0.024 & (2.296, 2.390) & 94.7
& 1.15 & 0.002 & (1.150, 1.156) & 95.6 \\
& Parametric with smaller misspecification
& 2.23 & 0.023 & (2.188, 2.279) & 95.5
& 1.15 & 0.002 & (1.146, 1.152) & 95.9 \\

\addlinespace[3pt]
\multirow{4}{*}{10,000}
& LSTM with sample splitting
& 2.01 & 0.011 & (1.990, 2.032) & 96.8
& 1.14 & 0.001 & (1.140, 1.143) & 97.0 \\
& LSTM without sample splitting
& 2.03 & 0.011 & (2.008, 2.052) & 96.1
& 1.14 & 0.001 & (1.142, 1.145) & 96.4 \\
& Parametric with larger misspecification
& 2.33 & 0.008 & (2.317, 2.347) & 77.1
& 1.15 & 0.000 & (1.151, 1.153) & 88.5 \\
& Parametric with smaller misspecification
& 2.23 & 0.007 & (2.217, 2.246) & 87.6
& 1.15 & 0.000 & (1.148, 1.150) & 92.2 \\

\bottomrule
\end{tabular}%
}
\end{table}





\begin{table}[!htbp]
\centering
\caption{\textbf{Complex longitudinal HIV data.} Risk differences and risk ratios relative to the natural course under three interventions.}
\label{tab:complex_rd_rr_all_interventions_coverage_truth}
\small
\setlength{\tabcolsep}{3pt}
\renewcommand{\arraystretch}{1.12}
\resizebox{\textwidth}{!}{%
\begin{tabular}{clcccc cccc}
\toprule
Sample size & Estimator
& \multicolumn{4}{c}{Risk difference ($\%$)}
& \multicolumn{4}{c}{Risk ratio} \\
\cmidrule(lr){3-6}\cmidrule(lr){7-10}
& & Estimate & Std.\ error & 95\% CI & Coverage ($\%$) & Estimate & Std.\ error & 95\% CI & Coverage ($\%$) \\
\hline
\multicolumn{10}{l}{\textbf{Always treat} (true risk difference = -4.00\%, true risk ratio = 0.76)} \\
\addlinespace[2pt]
\multirow{4}{*}{1,000}
& LSTM with sample splitting
& -4.79 & 0.035 & (-4.857, -4.719) & 83.9
& 0.72 & 0.002 & (0.719, 0.727) & 87.4 \\
& LSTM without sample splitting
& -4.32 & 0.034 & (-4.384, -4.250) & 88.7
& 0.74 & 0.002 & (0.738, 0.746) & 88.4 \\
& Parametric with larger misspecification
& -0.55 & 0.027 & (-0.606, -0.501) & 2.5
& 0.95 & 0.003 & (0.943, 0.953) & 32.7 \\
& Parametric with smaller misspecification
& -1.98 & 0.028 & (-2.031, -1.921) & 37.1
& 0.86 & 0.002 & (0.861, 0.868) & 60.7 \\

\addlinespace[3pt]
\multirow{4}{*}{10,000}
& LSTM with sample splitting
& -4.01 & 0.012 & (-4.037, -3.990) & 95.0
& 0.76 & 0.001 & (0.756, 0.759) & 94.5 \\
& LSTM without sample splitting
& -4.00 & 0.012 & (-4.028, -3.980) & 95.5
& 0.76 & 0.001 & (0.758, 0.761) & 94.8 \\
& Parametric with larger misspecification
& -0.55 & 0.008 & (-0.571, -0.538) & 0.0
& 0.95 & 0.001 & (0.946, 0.949) & 0.0 \\
& Parametric with smaller misspecification
& -2.00 & 0.009 & (-2.018, -1.983) & 0.0
& 0.86 & 0.001 & (0.863, 0.865) & 0.0 \\
\hline
\multicolumn{10}{l}{\textbf{Never treat} (true risk difference = 4.48\%, true risk ratio = 1.27)} \\
\addlinespace[2pt]
\multirow{4}{*}{1,000}
& LSTM with sample splitting
& 5.43 & 0.036 & (5.358, 5.501) & 82.0
& 1.32 & 0.002 & (1.311, 1.319) & 89.4 \\
& LSTM without sample splitting
& 4.85 & 0.035 & (4.785, 4.923) & 89.3
& 1.29 & 0.002 & (1.287, 1.295) & 92.1 \\
& Parametric with larger misspecification
& 0.63 & 0.026 & (0.576, 0.679) & 0.5
& 1.06 & 0.002 & (1.054, 1.064) & 21.3 \\
& Parametric with smaller misspecification
& 2.29 & 0.028 & (2.231, 2.342) & 29.0
& 1.16 & 0.002 & (1.153, 1.161) & 55.5 \\

\addlinespace[3pt]
\multirow{4}{*}{10,000}
& LSTM with sample splitting
& 4.50 & 0.012 & (4.472, 4.519) & 95.8
& 1.27 & 0.001 & (1.271, 1.273) & 96.1 \\
& LSTM without sample splitting
& 4.50 & 0.013 & (4.479, 4.530) & 95.0
& 1.27 & 0.001 & (1.269, 1.272) & 95.2 \\
& Parametric with larger misspecification
& 0.62 & 0.008 & (0.607, 0.639) & 0.0
& 1.06 & 0.001 & (1.057, 1.060) & 0.0 \\
& Parametric with smaller misspecification
& 2.31 & 0.009 & (2.295, 2.330) & 0.0
& 1.16 & 0.001 & (1.156, 1.158) & 0.0 \\
\hline
\multicolumn{10}{l}{\textbf{Dynamic intervention} (true risk difference = -3.12\%, true risk ratio = 0.81)} \\
\addlinespace[2pt]
\multirow{4}{*}{1,000}
& LSTM with sample splitting
& -3.45 & 0.023 & (-3.493, -3.402) & 91.8
& 0.80 & 0.001 & (0.798, 0.803) & 94.8 \\
& LSTM without sample splitting
& -3.24 & 0.022 & (-3.285, -3.200) & 94.6
& 0.81 & 0.001 & (0.804, 0.808) & 95.0 \\
& Parametric with larger misspecification
& -0.45 & 0.009 & (-0.472, -0.436) & 0.0
& 0.96 & 0.001 & (0.956, 0.959) & 0.5 \\
& Parametric with smaller misspecification
& -1.49 & 0.013 & (-1.519, -1.467) & 6.8
& 0.90 & 0.001 & (0.896, 0.899) & 19.5 \\

\addlinespace[3pt]
\multirow{4}{*}{10,000}
& LSTM with sample splitting
& -3.07 & 0.008 & (-3.085, -3.053) & 96.3
& 0.81 & 0.000 & (0.813, 0.815) & 97.8 \\
& LSTM without sample splitting
& -3.06 & 0.009 & (-3.073, -3.040) & 95.5
& 0.82 & 0.001 & (0.815, 0.817) & 97.1 \\
& Parametric with larger misspecification
& -0.45 & 0.003 & (-0.457, -0.445) & 0.0
& 0.96 & 0.000 & (0.957, 0.958) & 0.0 \\
& Parametric with smaller misspecification
& -1.51 & 0.004 & (-1.518, -1.500) & 0.0
& 0.90 & 0.000 & (0.897, 0.898) & 0.0 \\

\bottomrule
\end{tabular}%
}
\end{table}





\newpage

\clearpage

\end{document}